\documentclass{PoS}
\usepackage[mathscr]{euscript}
\usepackage{graphicx}
\usepackage{amsmath}
\usepackage{subfigure}
\graphicspath{{figures/}}

\def\Dslash{{D \!\!\!\!\!\!/}}
\def\chpt{\raise0.35ex\hbox{$\chi$}PT}
\def\subchpt{{\raisebox{1.2pt}{$\scriptstyle\chi$}{\rm PT}}}
\def\schpt{S\raise0.35ex\hbox{$\chi$}PT}
\def\rschpt{rS\raise0.35ex\hbox{$\chi$}PT}
\def\figref#1{Fig.~\ref{fig:#1}}
\def\Figref#1{Figure~\ref{fig:#1}}

\def\secref#1{Sec.~\ref{sec:#1}}

\def\Secref#1{Section~\ref{sec:#1}}

\def\gtwid{{\,\raise.3ex\hbox{$>$\kern-.75em\lower1ex\hbox{$\sim$}}\,}}
\def\ltwid{{\,\raise.3ex\hbox{$<$\kern-.75em\lower1ex\hbox{$\sim$}}\,}}

\def\ie{{\it i.e.},\ }
\def\eg{{\it e.g.},\ }
\def\et{{\it et al.}}

\def\cM{{\cal M}}

\def\cO{{\cal O}}

\def\rcite#1{Ref.~\cite{#1}}
\def\rcites#1{Refs.~\cite{#1}}
\def\Rcite#1{Reference~\cite{#1}}
\def\eqn#1{\label{eq:#1}}

\def\eq#1{Eq.~(\ref{eq:#1})}

\def\prd#1{Phys.\ Rev.\ D {\bf #1}}
\def\plb#1{Phys.\ Lett.\ {\bf #1B}}
\def\npb#1{Nucl.\ Phys.\ {\bf B#1}}

\def\seillac{Nucl.\ Phys.\ {\bf B} (Proc.\ Suppl.) {\bf 4} (1988)}
\title{Effective Field Theories and Lattice QCD}

\ShortTitle{Effective Field Theories and Lattice QCD}

\author{C.\ Bernard
\\
Department of Physics, Washington University,
St. Louis, MO 63130, USA\\
E-mail: \email{cb@lump.wustl.edu}
}
\abstract{I describe some of the many connections between lattice QCD and effective field theories, focusing
in particular on chiral effective theory, and, to a lesser extent, Symanzik effective theory.  I first discuss the ways in which effective theories have enabled and supported
lattice QCD calculations.  Particular attention is paid to the inclusion of discretization
errors, for a variety of lattice QCD actions, into chiral effective theory.  Several other
examples of the usefulness of
chiral perturbation theory, including the encoding of partial quenching and of twisted boundary conditions,
are also described.  In the second part of the talk, I turn to results from lattice QCD for the
low energy constants of the two- and three-flavor chiral theories. I concentrate here on mesonic
quantities,  but the dependence of the nucleon mass on the pion mass is also discussed.  Finally I describe some recent preliminary
lattice QCD calculations by the MILC Collaboration relating to the three-flavor chiral limit.}

\FullConference{The 8th International Workshop on Chiral Dynamics,\\
                June 29 -- July 3, 2015\\
                Pisa, Italy}

\begin{document}

\section{Introduction}
There is a close connection between lattice QCD and effective field theories (EFTs).  Lattice calculations require the use of EFTs at both a qualitative 
level, to understand the physics being simulated as well as the lattice artifacts introduced, and at a quantitative level, to extrapolate lattice results to physical values of parameters.  At the 
same time, the lattice makes possible the computation of the parameters of
 an EFT from the underlying, more fundamental, dynamics.  In the case of 
QCD and its low-energy EFT, chiral perturbation theory (\chpt) \cite{SW1979,Gasser:1983yg}, lattice calculations  
allow us to extract low energy constants and explore the 
structure of \chpt.  This is 
facilitated by the fact that lattice QCD simulations have free parameters to 
adjust (such as quark masses, and finite volume) that do not exist in Nature.

EFTs are a powerful tool to describe physics in some limited range of scales.
EFTs are useful when the fundamental theory is too difficult to handle (or is unknown).
Typically, an EFT is generated by  ``integrating out'' the
high-energy modes of a theory (those above a cutoff $\Lambda$). This leaves a non-local
theory, which in turn may be expanded in inverse powers of $\Lambda$  times local operators (an 
operator product expansion). The resulting local theory at low energy is the EFT of interest.
In rare cases (\eg  heavy quark effective theory), the steps can actually be carried out (perturbatively).  
Usually, however, we just imagine performing the above steps and use symmetries to constrain the EFT.


A key reason that EFTs are useful in lattice QCD is that they help control the extrapolations or
interpolations necessary to extract real-world results from lattice simulations, and thereby reduce
the corresponding systematic errors.  The systematic errors in lattice calculations inherently 
include: (1) {\it continuum extrapolation error} --- we need to take the lattice spacing $a$ to zero,
and (2) {\it finite volume error} --- we need to take the space and time extents of the lattice, $L$ and $T$, to infinity.  
In addition, there is often a {\it chiral extrapolation error}: for practical reasons, one may choose the light quark 
masses  $m_u$ and $m_d$ larger than in the real world, so 
one must extrapolate lattice results to the physical mass values.  
Even if near-physical values are chosen,  which is now possible, one needs to interpolate to the precise 
physical values (which can only be found {\it a posteriori}), leading to
a (presumably small) {\it  chiral interpolation error}.

The following is a brief outline of the rest of my talk.  \Secref{uses} describes the uses of EFTs in
lattice QCD. I focus in particular on \chpt\ because of its central role in lattice physics.
The inclusion of discretization errors (\secref{discretization}) and partial quenching effects in the chiral theory (\secref{PQ})
are 
discussed in detail. Some more recent applications are  described in \secref{other}. 
Besides \chpt, several other EFTs give crucial insight and information about lattice calculations.
I also have a good deal to say about Symanzik effective theory \cite{Symanzik:1983dc}, which
encodes the effects of discretization errors in continuum language.  
Another EFT that has played an important role is
heavy quark effective theory (HQET) \cite{HQET}; but due to limitations in space and time, I only  
touch upon it in passing.  

\Secref{payback} discusses the ``payback'': \chpt\ results from the lattice.  The emphasis
is on mesonic results, which are presented in \secref{mesons}.  I summarize what is known from lattice QCD 
about the values of some low energy constants (LECs) and the ``convergence'' of SU(2) \chpt,  and describe 
some interesting new lattice research.   I then briefly talk about nucleons in \secref{nucleons}.  \Secref{chiral-limit} discusses some
preliminary work on the three-flavor chiral limit and the convergence of SU(3) \chpt.

Finally, I discuss
future prospects, both of the uses of EFTs for the lattice, and on results from the lattice, in 
\secref{prospects}

\section{Uses of EFTs in Lattice QCD}
\label{sec:uses}

The chiral effective theory of QCD, \chpt, 
gives the functional form of the expansion of hadronic quantities (such as meson and baryon
masses) in terms of 
quark masses, with all dependence explicit.
This is exactly as needed for extrapolation of lattice data at heavier-than-physical quark masses
to the physical point.  
To my knowledge, the chiral theory was first used in this way in 1981 \cite{Hamber:1981zn} in order 
to extrapolate  the
pion mass as $M_\pi^2 \propto m_q$, with ($m_q$ the quark mass).%
\footnote{This does not require the whole apparatus of \chpt, but only the GMOR relation 
\cite{GellMann:1968rz}, which comes from the fact that the pion is a pseudo-Goldstone boson and from the PCAC hypothesis.}

\begin{figure}[t!]
  \centering
  {\includegraphics[width=0.48\linewidth]{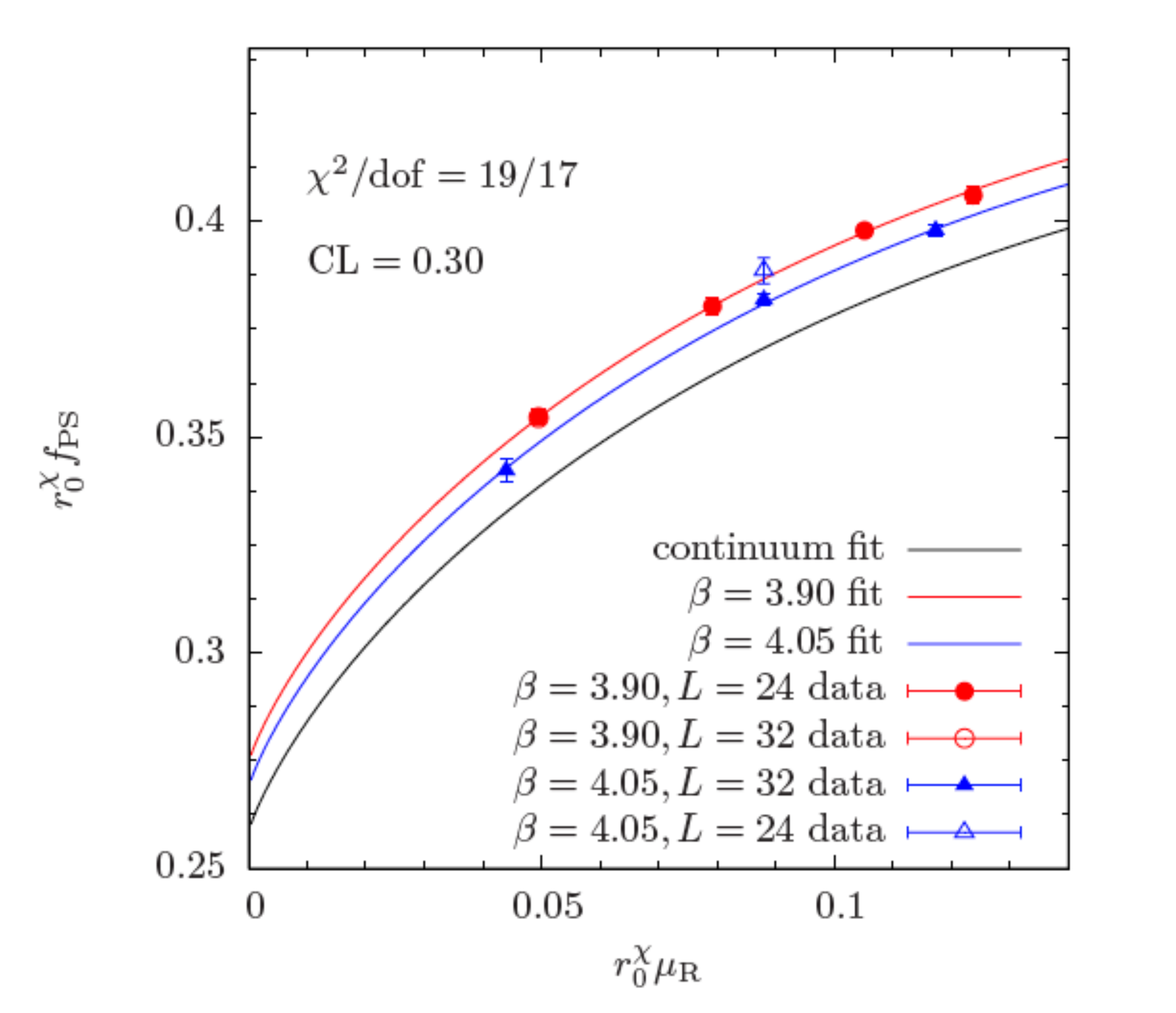}}
  \quad
  {\includegraphics[width=0.44\linewidth]{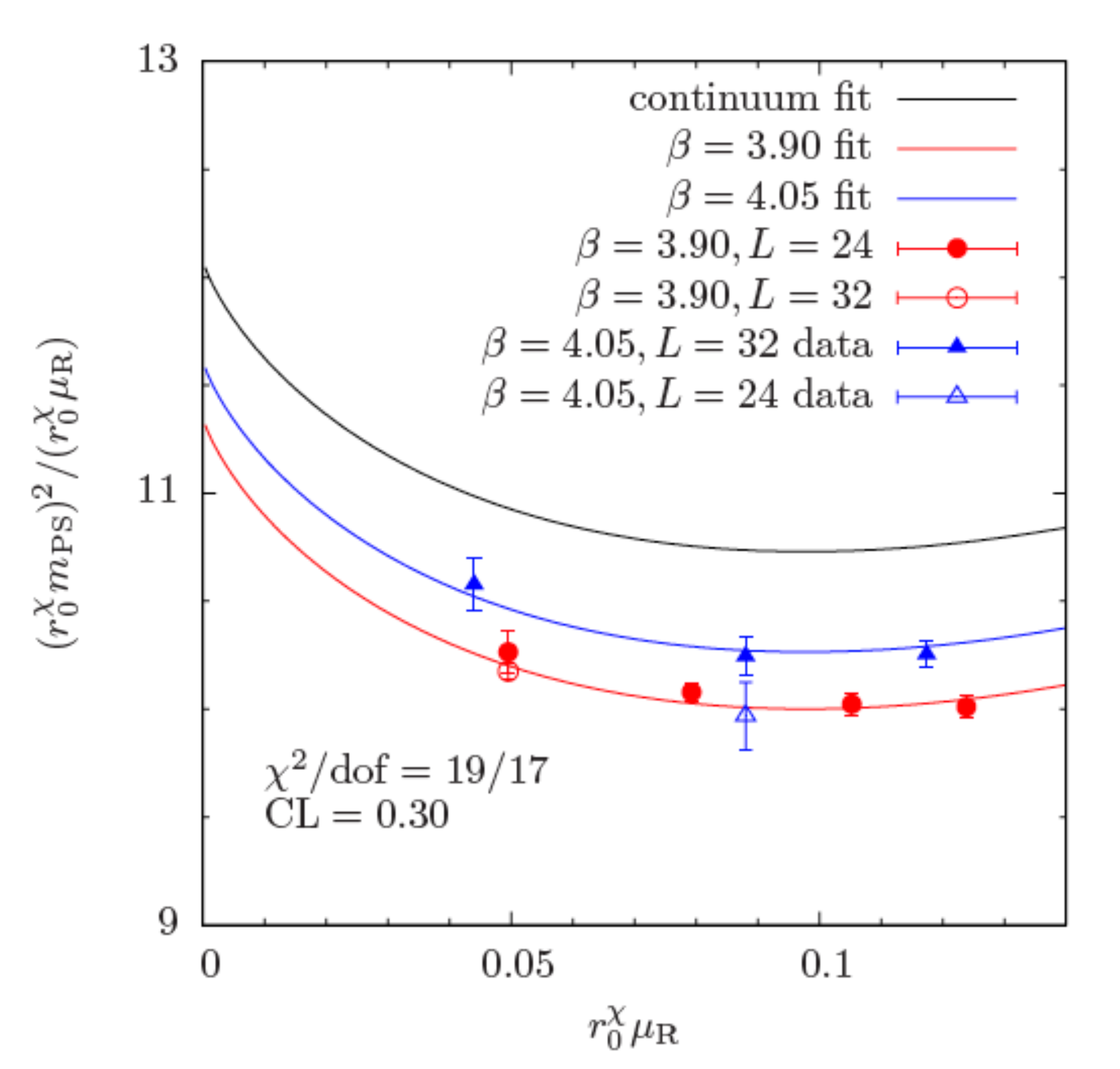}}\\
\vspace{-4mm}
\caption{\label{fig:fpi-ETM-2010} 
Chiral extrapolations of the pseudoscalar decay constant (left) and squared meson mass divided by quark mass (right), each plotted versus quark mass,
from the ETM Collaboration \cite{ETM2010}. Lattice data for two different lattice spacings
are shown: $a\approx0.08$ fm (red points) and $a\approx0.06$ fm (blue points). The black curves show
the result of extrapolating to the continuum.
}
\vspace{-0.08in}
\end{figure}

Chiral extrapolations of this kind remain the most common \chpt\ application to lattice calculations. \Figref{fpi-ETM-2010} 
shows extrapolations of $f_\pi$ and $M_\pi^2/\hat m$ by the ETM Collaboration \cite{ETM2010}
($\hat m$ is the average $u,d$ quark mass).
Discretization effects are fairly small, but clear: results for the different lattice spacings (in red and blue) differ
significantly. Thus,
as is essentially always the case, some $a$-dependence needs to be 
added to the continuum \chpt\ forms in order to fit lattice data. Here, the addition of simple analytic terms proportional to $a^2$ is sufficient. (We will see below
why terms of $\cO(a)$ do not appear.)

Later in the 1980's, it was realized \cite{FV} that \chpt\ also gives leading finite volume corrections, which come 
from pions looping around the finite volume, producing characteristic exponential ($e^{-M_\pi L}$) dependence.  Nowadays, 
the finite-volume dependence of lattice data is
commonly  fit and  extrapolated away using \chpt\ formulae.  The most recent FLAG review of lattice data \cite{Aoki:2013ldr}
requires $M_{\pi,{\rm  min}}L>4$ (or at least 3 volumes for fitting) to rate a lattice calculation as having good control of
finite volume errors.

\subsection{Including discretization errors in \chpt}
\label{sec:discretization}

In some cases (very precise lattice data, many degrees of freedom, larger discretization errors), the simple
$a$-dependence assumed in \figref{fpi-ETM-2010} may not be adequate to fit the lattice data.
Sharpe and Singleton \cite{Sharpe:1998xm} had the insight that \chpt\ can be modified to include lattice 
discretization errors.  This $a$-dependence is  related to the quark-mass dependence, so extrapolations may be
better controlled.  Non-polynomial terms in $a$ then arise from chiral loops.

To include  $a$-dependence in \chpt, we use, as an intermediate step, another EFT:  Symanzik effective theory (SET) 
\cite{Symanzik:1983dc}.
The SET takes the lattice-QCD theory at fixed lattice spacing $a$  as  ``fundamental,'' and then describes
that lattice theory at energy scales small compared to the cutoff:   $p\ll 1/a$.
At LO, the SET Lagrangian is just the continuum QCD Lagrangian.
Since $ap\ll 1$, one then needs to keep only low powers of $a$ as corrections.  To do this, we 
add on local operators with mass dimension greater than 4, multiplied by appropriate powers of $a$.
All local operators  allowed by the underlying lattice symmetries will appear.  Their coefficients are in principle computable
from the underlying lattice theory, but in practice are left as free parameters, similar to the LECs of \chpt.

As a first example, I describe the inclusion in \chpt\ of discretization errors from Wilson lattice quarks \cite{Wilson}. 
Wilson quarks remove lattice doublers by adding to the Lagrangian a term that breaks chiral symmetry (even for $m=0$), 
so the leading SET correction to the continuum theory is the Pauli operator: $a\;\bar q\sigma_{\mu\nu} G^{\mu\nu} q$.
Once the discretization effects are encoded in this way as one or more local operators, it is easy to include them 
in \chpt\ at low physical energies.
The method is standard {\it spurion}\/ approach that tells how the chiral-symmetry-breaking mass terms are represented in the
chiral Lagrangian.
For Wilson quarks, it is particularly simple, since the Pauli term and the mass term transform 
the same way under chiral symmetry:
\begin{equation}
{\cal L}^{\rm Wilson}_\subchpt =  \frac{f^2}{8}{\rm tr}(\partial_\mu \Sigma \partial_\mu\Sigma^\dagger) - \frac{Bf^2}{4}{\rm tr}(M\Sigma + M\Sigma^\dagger)
-a\,c_1{\rm tr}(\Sigma +\Sigma^\dagger)+a^2\,c_2{\rm tr}(\Sigma +\Sigma^\dagger)^2+\cdots
\eqn{Wilson-ChPT}
\end{equation}
Here the new LECs  $c_1$  and $c_2$ encode leading discretization effects in \chpt.
Sharpe and Singleton \cite{Sharpe:1998xm} showed from this \chpt\ that a lattice-artifact phase, the {\it Aoki phase}
\cite{Aoki}, was possible at fixed $a$ for very small $m$.  One reason this is important is that it
clarifies why massless pions appear as a function of the quark mass (and indeed at zero  quark mass in the
continuum limit) with two flavors of Wilson quarks, even though the action violates chiral symmetry. (I am not explaining the reasoning here; see \rcite{Sharpe:1998xm}.)

Twisted-mass quarks \cite{Frezzotti:2000nk} provide a second example. The theory starts with a doublet of
Wilson quarks, and adds a ``twist'' to the mass term:
\begin{equation}
\bar q\, (\Dslash+ m) q \ \ \to\ \   \bar q\, (\Dslash + m + i \mu \gamma_5 \tau_3) q.
\eqn{twisted-mass}
\end{equation}
In the continuum, the twist term proportional to $\mu$ can be rotated away by a non-singlet SU(2) 
chiral rotation, leaving only an untwisted mass $\!\sqrt{m^2+\mu^2}$.
But on the lattice, since the Wilson term (to remove doublers) is in the ``$m$ direction'' and therefore breaks this chiral 
symmetry, the twist has nontrivial
effects.
It provides two important improvements on ordinary Wilson quarks: (1) {\it Exceptional configurations} are eliminated.
Such configurations have statistical fluctuations from the Wilson term that cancel the normal mass term and
create numerical problems for Wilson-quark simulations.  (2) If $m$ is tuned to 0,  physical quantities have 
reduced discretization errors that start at $\cO(a^2)$, not $\cO(a)$  \cite{Frezzotti:2003ni}.  
The price of these improvements is violation of isospin symmetry at nonzero $a$.  

The isospin violation may be studied by adding the chiral representative of the
twist operator to the Wilson \chpt\ of Sharpe and Singleton, giving 
 twisted mass \chpt\ \cite{Munster:2004dj}.   The \chpt\ shows that the mass of the
$\pi^0$ is split from that of the $\pi^\pm$ by a term of $\cO(a^2)$.  This is verified by simulations;
see for example Fig.~4 of \rcite{ETM2010}.

The chiral theory for staggered quarks is my final example.  Staggered quarks have incomplete reduction of 
lattice doubling symmetry, resulting in an extra (unwanted) degree of freedom known as {\it taste}.
Each flavor of quark comes in 4 tastes.  Since  taste is unphysical, it  needs to be removed from the sea
by the simulation algorithm.  This is done with the {\it fourth root procedure} --- taking the fourth root
of the fermion determinant.%
\footnote{The justification of this procedure at the non-perturbative level is a subtle issue
that is outside the scope of this talk, but let  me just say that the staggered version of \chpt\ is helpful in understanding and taming 
the issue \cite{Bernard:2006zw}.}

Staggered SU(4) taste symmetry is exact in the continuum, but is violated on the lattice at $\cO(a^2)$ due
to the exchange of high-momentum gluons between quarks.
The SET is:
\begin{equation}
{\cal L}^{\rm stag}_{\rm SET} = \frac{1}{4} G^{\mu\nu}G^{\mu\nu} + \bar q\, (\Dslash + m) q 
+a^2\,\bar q (\gamma_\mu \otimes \xi_5) q\;\bar q (\gamma_\mu \otimes \xi_5) q + \cdots,
\eqn{staggered-SET}
\end{equation}
where the first two terms just constitute the continuum Lagrangian, and the 4-quark, $\cO(a^2)$, terms 
(of which only one is shown) come from gluon exchange and violate taste symmetry due to the presence of fixed taste matrices (here,
$\xi_5$).
The  staggered \chpt\  Lagrangian may then be derived  \cite{Lee:1999zxa,Aubin:2003mg} with the spurion 
method, and is
\begin{equation}
{\cal L}^{\rm stag}_{\subchpt} =  \frac{f^2}{8}{\rm tr}(\partial_\mu \Sigma \partial_\mu\Sigma^\dagger) - 
\frac{Bf^2}{4}{\rm 
tr}(M\Sigma + M\Sigma^\dagger)
-a^2\,C_1{\rm tr}(\xi_5\Sigma \xi_5\Sigma^\dagger)+\cdots,
\eqn{staggered-ChPT}
\end{equation}
where $\Sigma$ is now a $4n \times 4n$ matrix, with $n$ the number of flavors, 
and $C_1$ is a new LEC that
depends on the particular implementation of staggered fermions.  This is unlike the case of
the familiar  continuum LECs, which should be
independent of the lattice action, up to possible discretization errors from higher-dimension operators in the SET that do not 
break any continuum symmetries.

\begin{figure}[t!]
  \centering
  {\includegraphics[width=0.48\linewidth]{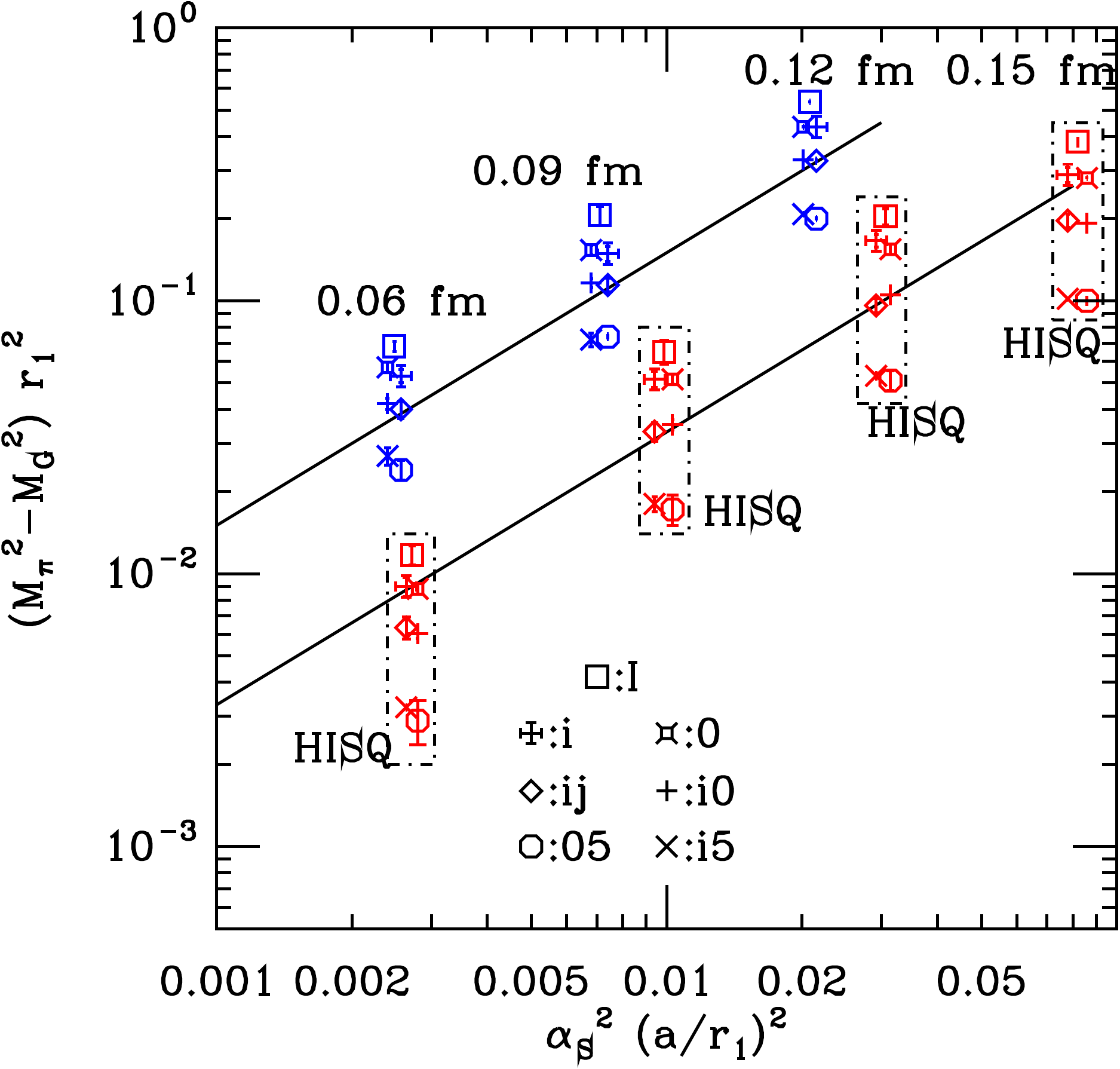}}\\
\vspace{-2mm}
\caption{\label{fig:staggered-splittings} 
Pion taste splittings from \rcite{Bazavov:2012xda} versus $(\alpha_S a)^2$ for two versions of staggered quarks:   asqtad (blue)  and the newer, more highly improved version, HISQ \cite{Follana:2006rc} (red).  The diagonal lines have slope 1, and show  what is expected if the splittings
are linear in $(\alpha_S a)^2$.
}
\vspace{-0.08in}
\end{figure}

Expanding ${\cal L}^{\rm stag}_\subchpt$ to quadratic order, one finds as expected 16 pions for each (non-singlet) flavor combination.
At fixed lattice spacing, only one of the pions for each flavor combination is a true Goldstone boson whose mass vanishes in
the chiral limit; this pion is
the one associated with the one non-singlet chiral symmetry unbroken by discretization effects.  The other 15 pions are
raised above the Goldstone one by $\cO(a^2)$ terms  (times $\alpha_S$, or  $\alpha^2_S$ if the staggered action is
improved).   \Figref{staggered-splittings} shows the splittings between pions of various tastes and the Goldstone one
as a function of $(\alpha_S a)^2$.  The picture is exactly as predicted by staggered \chpt, with these splittings vanishing
linearly with $(\alpha_S a)^2$ as the continuum limit is approached and taste symmetry is restored.  With the more highly
improved HISQ action the splittings are smaller, implying that LECs such as $C_1$ are smaller.

\subsection{Partial Quenching}
\label{sec:PQ}
So far, I have described how including discretization effects in \chpt\  can help elucidate lattice artifacts 
such as the Aoki phase (Wilson quarks),
pion isospin-violations (twisted mass quarks), or 
pion taste-splittings (staggered quarks).
But ultimately the goal is to extract physical results, so a key use of $a$-dependent \chpt\ is to
guide continuum extrapolations.  The idea is to fit quark-mass dependence and lattice-spacing dependence 
together, using expressions from the appropriate \chpt, and thereby
to reduce systematic errors.
Such fits are often done in a {\it partially quenched} context:  we may choose the valence quarks to have  
different masses from those of the sea quarks.
This is useful because valence quarks are cheap compared to sea quarks.  The computational resources 
necessary to generate a gluon configuration, including the back effect of sea quarks (the fermion
determinant), are large. It therefore makes sense to extract as much information as possible from a given 
configuration by the inexpensive  calculation of several valence quark propagators, with a range of masses, 
in that background gluon configuration. This is called partially quenching because valence quarks are 
quenched (forbidden from appearing in virtual loops), but sea quarks are not quenched.
To understand partial quenching at the Lagrangian level, we imagine adding ghost (bosonic!) quarks, with the same mass matrix as the valence quarks, to cancel the virtual loops (determinant) of the valence quarks \cite{Morel:1987xk}.

The partially quenched QCD (PQQCD) Lagrangian in the continuum is
\begin{equation}
{\cal L}_{\rm PQQCD} = \frac{1}{4} G^{\mu\nu}G^{\mu\nu} + \bar q\, (\Dslash + {\cal M}) q\; +\bar {\hat q}\, (\Dslash + {\cal M'}) \hat q + \bar {\tilde q}\, (\Dslash + {\cal M'}) \tilde q,
\eqn{PQQCD}
\end{equation}
where $q$ is a vector of sea quarks (with mass matrix $\cM$), while $\hat q$ and $\tilde q$ are the
valence quarks and ghost quarks, respectively (both with mass matrix $\cM'$).
When  $\cM'=\cM,\;$   PQQCD reduces to QCD.%
\footnote{More precisely, in this limit the QCD Green's functions and physical quantities are a proper subset of those in PQQCD.}

Then the Lagrangian for partially quenched \chpt\  (PQ\chpt) at LO is \cite{Bernard:1993sv}

\begin{equation}
{\cal L}_{\rm PQ\subchpt} = \frac{f^2}{8}{\rm str}(\partial_\mu \Sigma \partial_\mu\Sigma^\dagger) - \frac{Bf^2}{4}{\rm str}(M\Sigma + M\Sigma^\dagger).
\eqn{PQChPT}
\end{equation}
This looks fairly standard, but  $\Sigma$  is a $(n_{\rm sea}+2n_{\rm val})\times(n_{\rm sea}+2n_{\rm val})$ matrix, with pions 
of all combinations of quarks (sea-sea, sea-valence, sea-ghost, valence-valence, ...),  and 
the chiral symmetry group is the {\it graded group} \cite{DeWitt}  
${\rm SU}(n_{\rm sea}+n_{\rm val} | n_{\rm val}) \times {\rm SU}(n_{\rm sea}+n_{\rm val} | n_{\rm val})$ instead of the usual  
${\rm SU}(n_{\rm sea}) \times {\rm SU}(n_{\rm sea})$.  A graded group has some Grassmann generators, because 
some transformations take fermions into bosons and {\it vice versa}, as in supersymmetry.
Similarly, str in \eq{PQChPT} is 
the {\it supertrace}, which has some minus signs (in the fermion-fermion block of the matrix) relative to a 
normal trace.
The mass matrix $M= {\rm diag}(\cM,\cM',\cM')$.  I emphasize here that the ghosts have been introduced 
only as a theoretical tool to
understand partial quenching in \chpt.  In simulations, we may set, by hand,  the masses
of valence quarks (in quark propagators) different from the those of sea quarks  (in the quark determinant); no ghosts
are required.


In the loop expansion, PQ\chpt\ calculations  are just like those for standard \chpt, except that
the fermionic mesons (sea-ghost or valence-ghost pions) introduce minus signs in loops. 
That serves to cancel the unwanted loops associated with valence quarks.
Since valence particles on external lines do not appear in loops, it is clear that PQQCD violates unitarity;
alternatively, it is enough to note that the Lagrangian contains spin-1/2 bosons.
Unitarity is restored in the limit when valence and sea masses are equal.%
\footnote{For all physical correlation functions corresponding to those in ordinary QCD.}
However, even for unequal valence and sea masses, the LECs of PQ\chpt\ are the same
as those of the real world, since LECs do not depend on quark masses \cite{Sharpe:2000bc}.  This
is the main reason why PQQCD and PQ\chpt\ are useful.

One may wonder whether PQ\chpt\ is really justified as an effective field theory. 
The original justification for ordinary \chpt\ by Weinberg \cite{SW1979} used 
analyticity, clustering, and unitarity, but here unitarity is violated.  Recently, Maarten Golterman
and I revisited the issue to try to put PQ\chpt\ on a firmer footing \cite{Bernard:2013kwa}.
The argument is based on Leutwyler's justification for \chpt\ \cite{HLfound}, which uses 
clustering and locality, but not unitarity.
Locality and clustering  are properties that can exist for a Euclidean field theory 
even if it is not unitary.
We showed that Euclidean PQQCD on the lattice
has a transfer matrix and hence a Hamiltonian.
The Hamiltonian is not Hermitian, but has a positive definite real part for nonzero quark mass.
From that, we were able to argue (modulo some ``mild'' assumptions) that the theory clusters, and then
 that PQ\chpt\ is indeed the correct EFT for PQQCD.

\begin{figure}[t!]
  \centering
  {\includegraphics[width=0.48\linewidth]{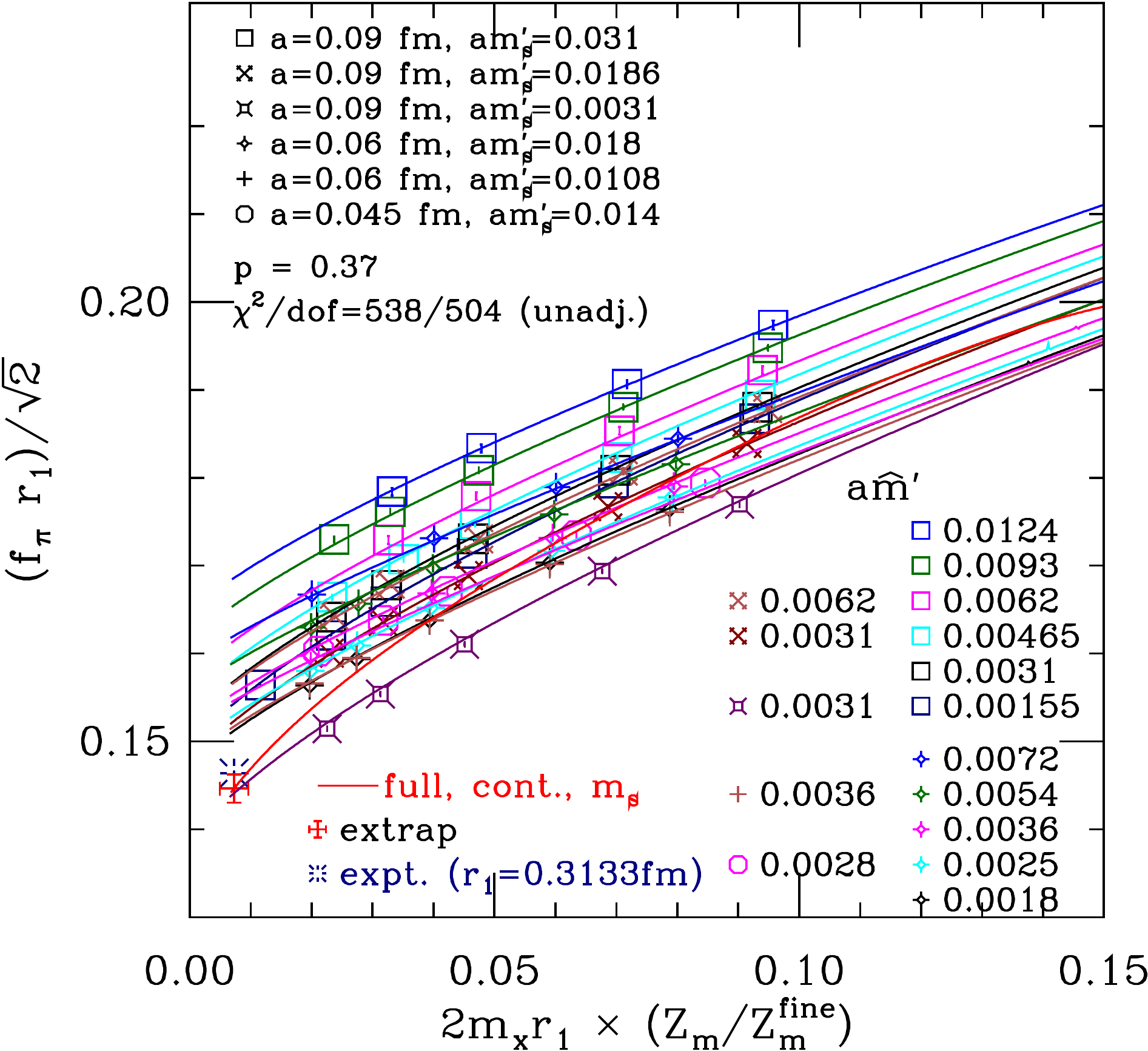}}
  \quad
  {\includegraphics[width=0.44\linewidth]{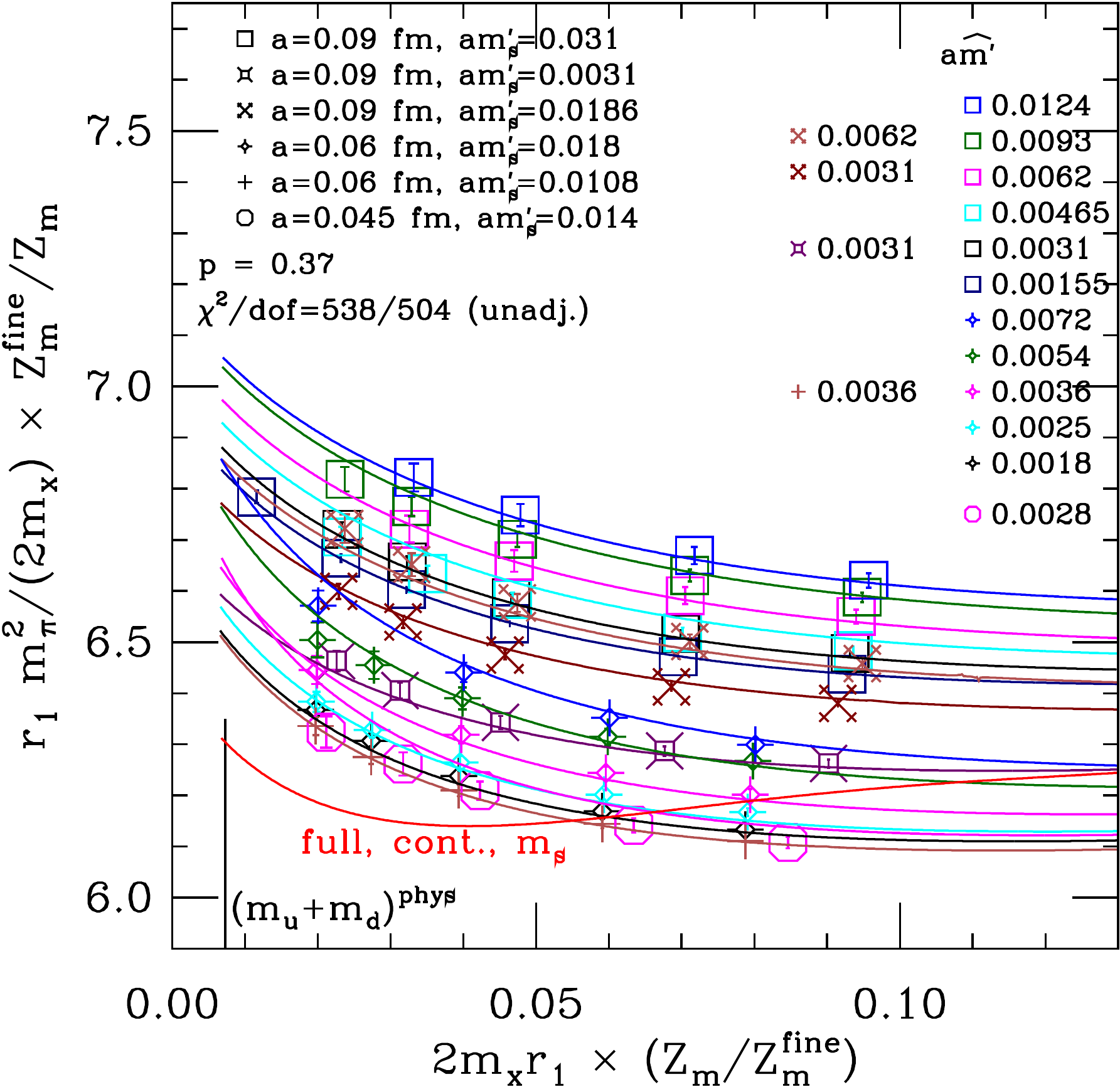}}\\
\vspace{-2mm}
\caption{\label{fig:MILC-Lat10} 
Continuum and chiral extrapolation of partially quenched staggered lattice data 
for decay constants (left) and squared meson mass divided by quark mass (right), each plotted versus quark mass, from the MILC Collaboration \cite{MILC-Lat10}.  Sets of points with the same color and 
shape represent fixed $a$ and sea-quark masses, with changing valence-quark mass. 
}
\vspace{-0.08in}
\end{figure}

Fitting partially quenched lattice data with a PQ\chpt\ that also includes discretization effects 
gives good control of lattice errors when  extrapolating to 
physical quark masses and to the continuum and unitary limits.
\Figref{MILC-Lat10} shows such a fit of lattice data for decay constants and masses of pseudoscalar
mesons.  The fit lines of various colors show the effect of varying valence quark mass with
fixed lattice spacing and sea-quark masses.  
The red lines
represent the results as a function of valence mass after extrapolating to the continuum and
taking the unitary limit (sea mass set equal to valence mass).  The change in slope between
the fit lines and the red extrapolated line shows the significant effect of partial quenching,
which is well accounted for by the PQ\chpt.  The final extrapolation to physical valence
quark mass gives results with $\sim\!1\%$ errors 
even though the lattice data typically has $\sim\!10\%$ discretization, mass, or 
partial-quenching corrections. This shows the power of an EFT approach that includes the effects
of lattice artifacts.
 
\subsection{Some Other Recent Applications of EFTs}
\label{sec:other}

The inclusion of  discretization errors and partial quenching effects in \chpt, which have been
discussed so far, are important for lattice calculations, but not new. I now briefly discuss
three recent, or ongoing, developments in the application of EFTs to lattice QCD.

The first deals with what are called {\it twisted boundary conditions}.
With periodic boundary conditions, lattice momenta are limited to values 
$p = 2\pi n / L$, where $L$ is the spatial lattice dimension, and $n$ is an integer.
Even with the large volumes used in many modern simulations, $L\sim 5$ fm, momenta are spaced 
by $\sim\!250$ MeV, making it difficult to map out the momentum dependence of quantities
such as semileptonic form factors or structure functions.
A solution \cite{Bedaque:2004kc,deDivitiis:2004kq}
to this problem is to  give (some) quarks twisted boundary conditions:
$q(x+L) = e^{i\theta} q(x)$.  
Then the allowed momenta are $p = (2\pi n + \theta) / L$, and $\theta$ can be adjusted to
give any desired momentum. 

The finite-volume effects are different with twisted boundary conditions than with periodic ones.
The differences can be worked out in \chpt.  The first attempt to do so was by
Sachrajda and Villadoro \cite{Sachrajda:2004mi}.  However, certain contributions
that vanish with periodic boundary conditions by (lattice) rotational symmetries
were omitted in \rcite{Sachrajda:2004mi}, as pointed out by Jiang and Tiburzi 
\cite{Jiang:2006gna}.  The issue has recently been further clarified by  Bijnens and Relefors
\cite{Bijnens:2014yya}, who systematically work out the finite-volume corrections for masses,
decay constants, and electromagnetic form factors.
The infinite volume integral
\begin{equation}
\int \frac{d^4\, k}{2\pi^4} \frac{k_\mu}{k^2 +m^2} =0,
\eqn{infvol-integral}
\end{equation}
which vanishes because it is odd under $k\to-k$ (or, because of rotational symmetry), provides
the simplest example.  With twisted boundary conditions in finite volume, the integral becomes
\begin{equation}
\frac{1}{L^4}\;\sum_{k_\nu = \frac{2\pi n_\nu +\theta_\nu}{L}} \frac{k_\mu}{k^2 +m^2} \not=0,
\eqn{twisted-sum}
\end{equation}
which is nonzero because the twist breaks the $k\to-k$ symmetry. (I have taken $T=L$ for simplicity.)

A second recent application of EFTs is the development of
\chpt\ for heavy-light mesons (D or D$_{\rm s}$), including discretization errors, when 
both heavy and light quarks are staggered \cite{Komijani:2012fq}.
The HISQ version of staggered quarks \cite{Follana:2006rc} makes possible charm-quark simulations 
with the staggered action. This is a
highly improved action, which, by removing the leading discretization effects due to
the charm quark (eliminating all tree-level $(am_c)^2$ terms and the major tree-level $(am_c)^4$ effects), allows us to treat the charm mass as small in lattice units.  Thus we may assume
$a m_c \ll 1$  (even though we may in fact have $a m_c \ltwid 1$). 

From the point of view
of the current talk, the \chpt\ developed in \rcite{Komijani:2012fq} is interesting because it
employs a chain of three EFTs. First, the lattice theory is replaced with a SET using the fact that
the light quark and charm quark masses obey, effectively, $am_\ell \ll 1$, $a m_c \ll 1$.
Then, assuming $\Lambda_{QCD}/m_c \ll 1$, the charm quark mass scale  may be removed from the
theory by treating the charm quark with HQET. The last step is to take $m_\ell/\Lambda_{QCD}\ll 1$
and treat the light quark degrees
of freedom in \chpt.  The resulting chiral theory goes by the long-winded name of
{\it heavy-meson, rooted, all-staggered chiral perturbation theory}
(HMrAS\chpt).

\begin{figure}[t!]
  \centering
  {\includegraphics[width=0.9\linewidth]{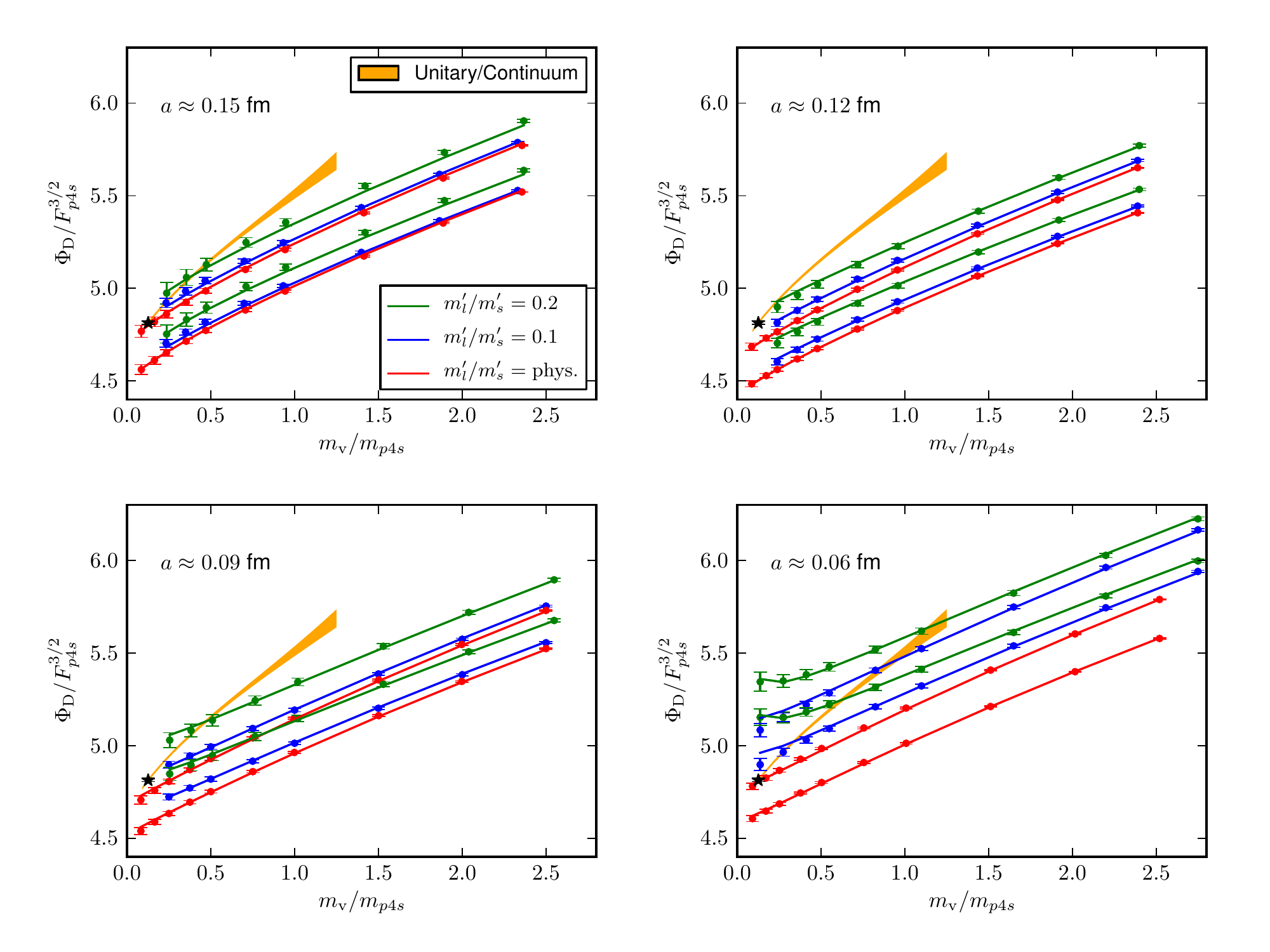}}\\
\vspace{-5mm}
\caption{\label{fig:MILC-fD} 
Fit to lattice data for heavy-light decay constants (more
precisely $\Phi_{D} = f_{D}\sqrt{M_D}$), as a function of valence quark mass,
for four lattice spacings, from $a\approx0.15$ fm (upper left) to $a\approx0.06$ fm (lower right)
\cite{Bazavov:2014wgs}.
Different colors indicate different values of the light sea-quark mass, with red corresponding
to an approximately physical mass value.  The two lines of each color in each plot are for
two values of the charm valence quark mass.  The orange band gives the result after extrapolating
to the continuum and to infinite volume, and setting the light sea mass equal to the light valence mass; the black star shows
the physical result for $\Phi_{D^+}$.
}
\vspace{-0.08in}
\end{figure}

\Figref{MILC-fD} shows a fit of lattice data for heavy-light 
decay constants generated by the Fermilab/MILC collaboration 
\cite{Bazavov:2014wgs} to the functional form derived in HMrAS$\chi$PT.
Some of the data is generated at approximately physical values
of the sea quark masses (2+1+1 flavors).  Staggered discretization errors and 
the effects of partial quenching are included in the fit.
The fit is able to reproduce the features of the large data set (366 points), including
all correlations, and has $\chi^2/{\rm dof}=347/339$, with $p\!=\!0.36$.
After setting sea and valence masses equal, extrapolating to the continuum, and
interpolating to the physical values of the valence quark masses, the result is \cite{Bazavov:2014wgs}
\begin{eqnarray}
f_{D^+}&=&212.6(0.4)(^{+1.0}_{-1.2})\;{\rm MeV}\nonumber\\
f_{D_s}&=&249.0(0.3)(^{+1.1}_{-1.5})\;{\rm MeV}.
\eqn{MILC-fD}
\end{eqnarray}
The errors are 
at the sub-percent level of precision, approximately 2 to 4 times (depending on the quantity)
 more precise than the previous most-precise lattice calculations.

Finally, I mention here the application of \chpt\ to observables defined using
gradient flow, the gauge-covariant smoothing
of gluon \cite{Luscher:2009eq,Narayanan:2006rf} and quark \cite{Luscher:2013cpa} fields.  
Smoothing eliminates or reduces short-distance divergences, making expectation values and correlators constructed from smoothed fields finite or multiplicatively renormalizable
\cite{Luscher:2010iy,Luscher:2011bx}. This means that such observables have many practical uses.     The
most important application, at least at present, is fixing the lattice scale from
the smoothed gluon condensate \cite{Luscher:2010iy,Borsanyi:2012zs}.  The main advantages over most alternative
scale-setting procedures are that gradient flows scales have small statistical errors, weak and calculable dependence on sea-quark masses, 
and no dependence on valence quarks at all.  

The sea-quark mass dependence of gradient-flow scales has been worked out in \chpt\ through one-loop 
by B\"ar and Golterman \cite{Bar:2013ora}.  Since the gluonic energy density is a chiral invariant, its expansion in
terms of the chiral field $\Sigma$ and the masses is
very much like that of the QCD Lagrangian itself, but with the flow-time dependence encoded in the LECs.  
An important point is   that a constant (independent of the chiral field and
masses), which is always dropped in the case of the Lagrangian, is relevant here and represents the energy density in the chiral limit.
Thus LO for the energy density is $\cO(p^0)$ in the usual chiral counting, and $\cO(p^2)$, which gives linear quark-mass
dependence, is already NLO.   Chiral logarithms, generated at one loop, are NNLO. Fitting lattice data
for gradient flow scales to the NNLO formulae from \rcite{Bar:2013ora} therefore allows one to
reduce already small sea-quark effects even further.  This been done in \rcite{Bazavov:2015yea}.
I note that \rcite{Bar:2013ora} also computes the chiral condensate and decay constant in \chpt\ at nonzero flow time.  In
the chiral limit, the results reduce to those previously calculated by L\"uscher \cite{Luscher:2013cpa}.

\section{\chpt\ Results from the Lattice}
\label{sec:payback}

In principle, lattice techniques make possible the computation of LECs of the effective theory from fundamental QCD.
In practice, this is easiest for LECs affecting pseudoscalar meson masses and leptonic decay constants.
These can be calculated from the quark-mass dependence of 2-point, connected, Euclidean correlation functions, which
are relatively simple quantities for lattice QCD to treat.  
This makes lattice QCD a nice complement to experiments, which give little constraint on quark-mass dependence 
since quark masses are fixed in Nature.
On the other hand, LECs affecting momentum dependence of hadronic scattering amplitudes are just the opposite: relatively
straightforward for experiments, but difficult on the lattice.  They not only require computation of $n$-point functions,
but, more importantly, require  extraction of excited-state amplitudes in finite volume \cite{Luscher:1990ux}
because the ground states in these Euclidean correlators are the states with momentum just at  threshold
\cite{Maiani:1990ca}.

\subsection{Results for Pseudoscalar Mesons}
\label{sec:mesons}
\begin{figure}[t!]
  \centering
  {\includegraphics[width=0.48\linewidth]{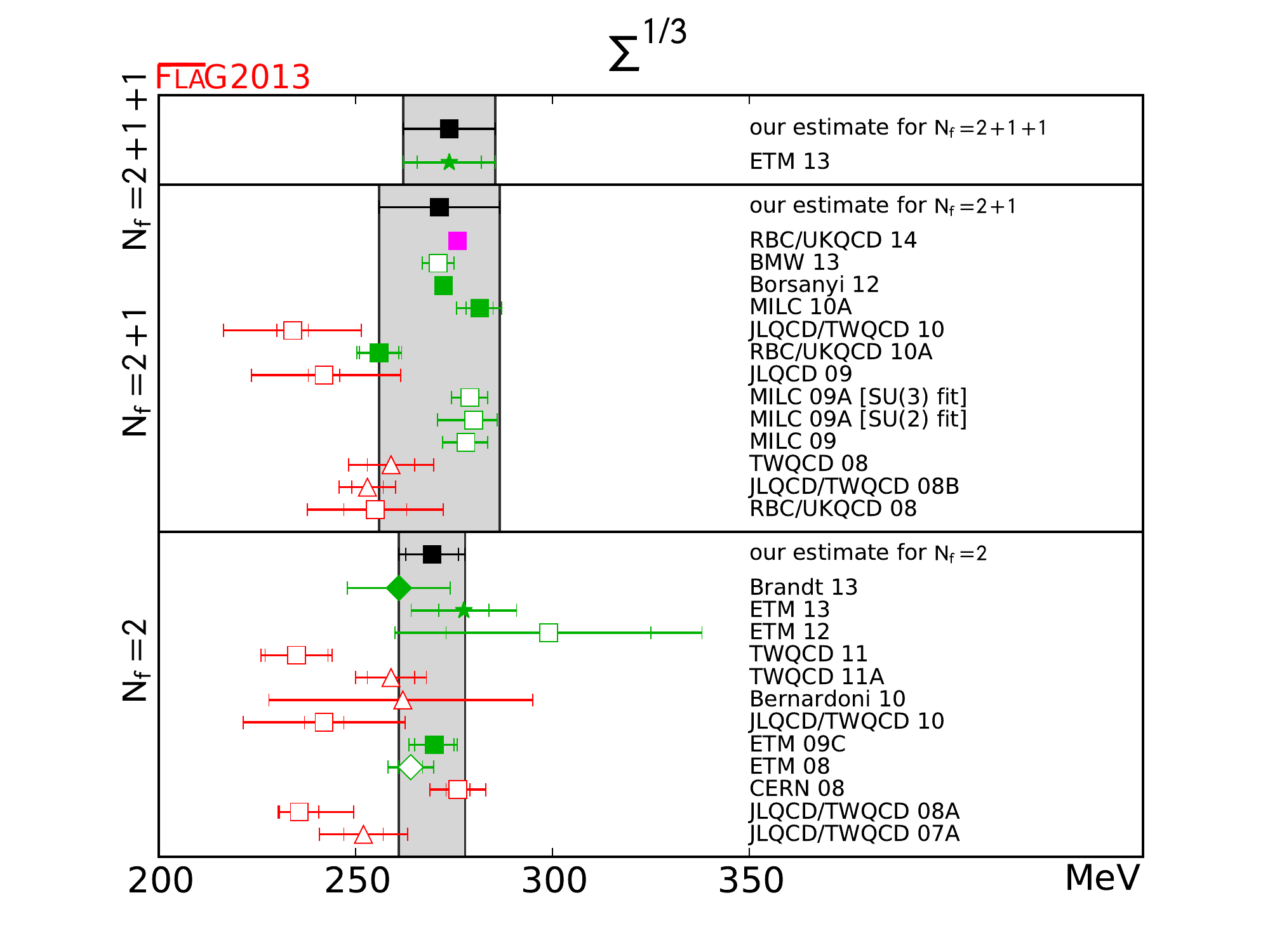}}
  \quad
  {\includegraphics[width=0.48\linewidth]{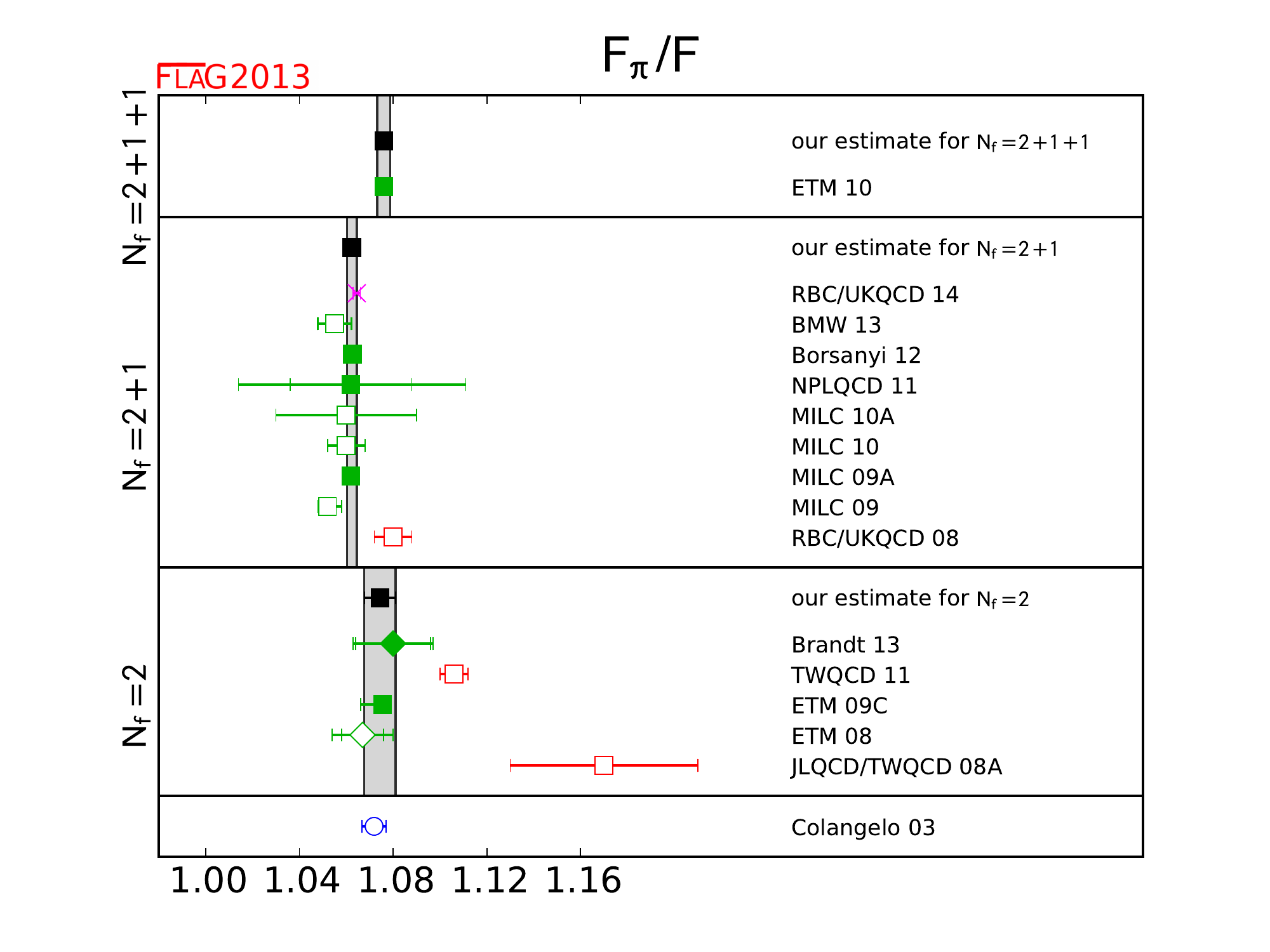}}\\
\vspace{-4mm}
\caption{\label{fig:su2-Sigma-F}
FLAG summary plots \cite{Aoki:2013ldr} from 2013 for the quark condensate (left) and $F_\pi/F$ (right) in the two-flavor
chiral limit.  The magenta points, from RBC/UKQCD$\,$14 \cite{Blum:2014tka}, have been added by me; for  $F_\pi/F$, I use
a cross instead of a square because the errors bars are significantly smaller than the size of a square.
}
\vspace{-0.08in}
\end{figure}

\Figref{su2-Sigma-F} shows the most recent (2013) FLAG \cite{Aoki:2013ldr} summary plots  for the quark condensate
$\Sigma=|\langle\bar u u\rangle|$ and decay constant $F$ in the two flavor chiral limit: $m_u,m_d\to0$. I have superposed
recent results from the RBC/UKQCD Collaboration \cite{Blum:2014tka} in magenta.  The 2013 FLAG averages in the case of
$N_f=2+1$ are
\begin{equation}
\Sigma(\mu=2\,{\rm GeV})= (271(15)\; {\rm MeV})^3,\qquad \frac{F_\pi}{F} = 1.0624(21).
\eqn{su2-Sigma-F}
\end{equation}
The small errors on the RBC/UKQCD$\,$14 results show how much the field has progressed in just the past two years.  As 
noted by FLAG, the non-monotonic behavior of $F_\pi/F$ as the number of flavors changes suggests that the systematic
errors of one or more calculations may be underestimated.  Since the $N_f=2+1$ simulations
of both RBC/UKQCD$\,$14 and Borsanyi$\,$12 include ensembles with physical light ($u,d$) quark masses, whereas the 
$N_f=2+1+1$ simulations of
ETM$\,$10 do not, I suspect the errors of the ETM$\,$10 calculation may be somewhat underestimated due to the difficultly in 
controlling the extrapolation from their lightest masses ($M_\pi\approx 270$ MeV) down to the chiral limit.  Of course, there
are other possibilities, and, in particular, there is no theorem that states that the dependence on the number of flavors must be
monotonic. 

\begin{figure}[t!]
  \centering
  {\includegraphics[width=0.6\linewidth]{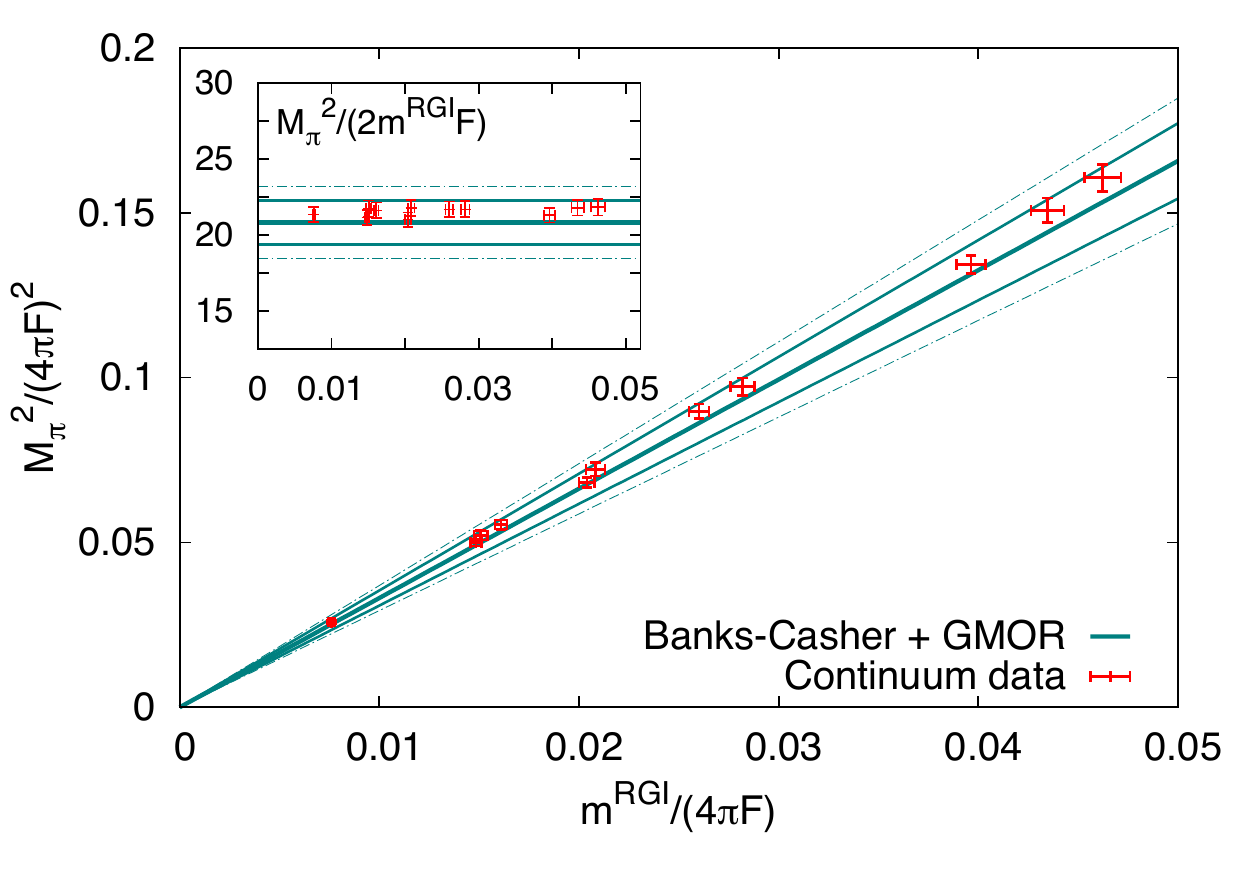}}\\
\vspace{-6mm}
\caption{\label{fig:Engel}
$M_\pi^2/(4\pi F)^2$ versus quark mass from  \rcite{Engel:2014cka}.  The green lines show the slope from
the condensate, and the red points come from direct lattice determinations of $M_\pi$, 
after extrapolation to the continuum.  
}
\vspace{-0.03in}
\end{figure}

A recent calculation of the quark condensate in the two-flavor chiral limit is presented by 
Engel {\it et al.}\ \cite{Engel:2014cka}.  
They first calculate the condensate from the eigenvalue density, {\it a la} Banks-Casher \cite{Banks:1979yr}.
From the GMOR relation \cite{GellMann:1968rz}, the condensate gives the slope of  $M_\pi^2$ with quark mass.
\Figref{Engel} shows the slope determined from the eigenvalue density as the dark green line, with lighter green lines 
indicating the errors.  The red crosses come from the direct lattice data for $M_\pi^2$, extrapolated to the continuum.
The agreement  is excellent.  The calculation is a  nice demonstration of the deep structure of chiral symmetry
breaking in QCD.  Engel {\it et al.}\ find $\Sigma = (263(3)(4)\; {\rm MeV})^3$ with $N_f=2$ at renormalization scale 2 GeV;
the corresponding FLAG 2013 average is $\Sigma = (269(8)\; {\rm MeV})^3$.

\begin{figure}[t!]
  \centering
  {\includegraphics[width=0.48\linewidth]{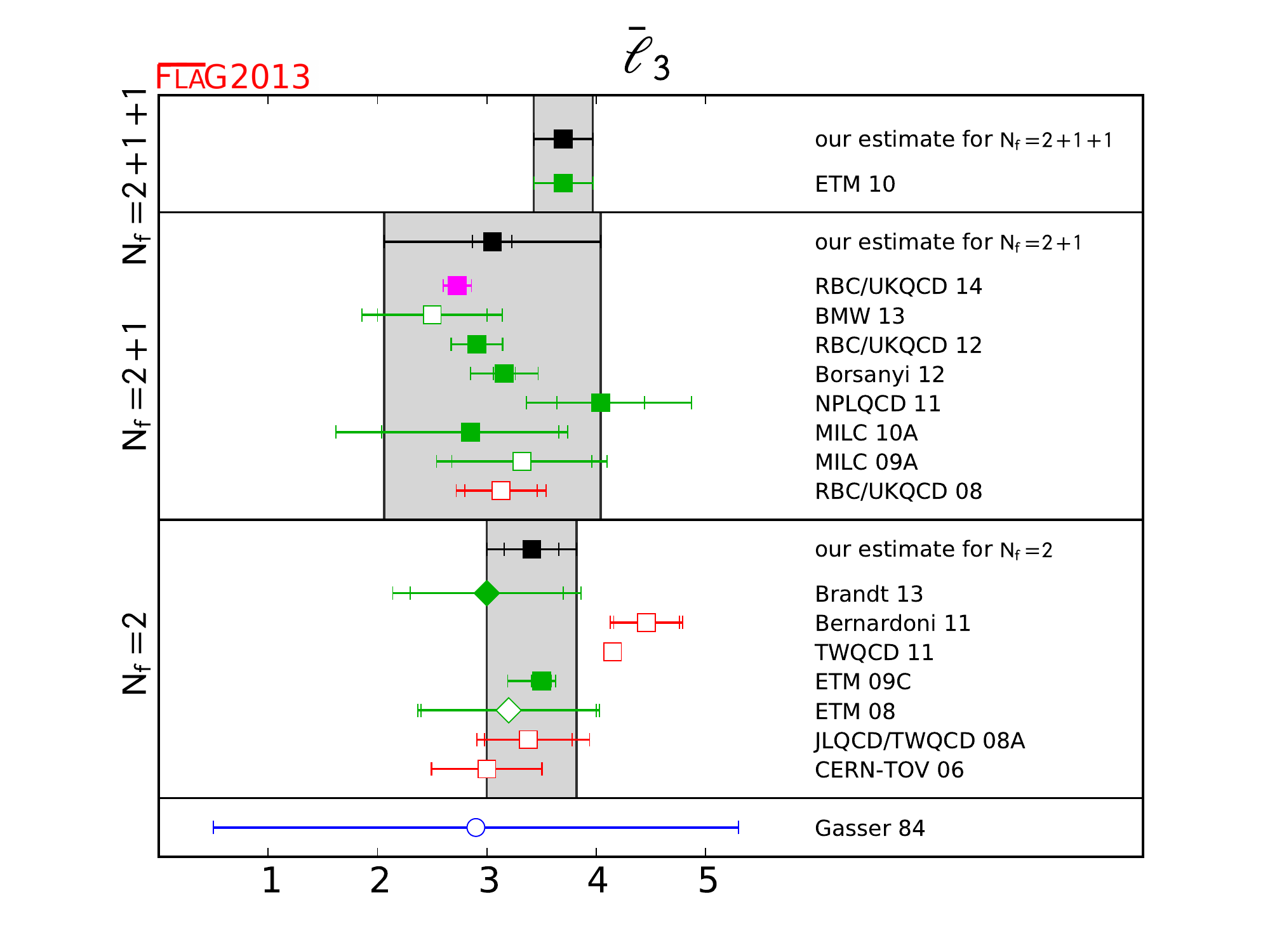}}
  \quad
  {\includegraphics[width=0.48\linewidth]{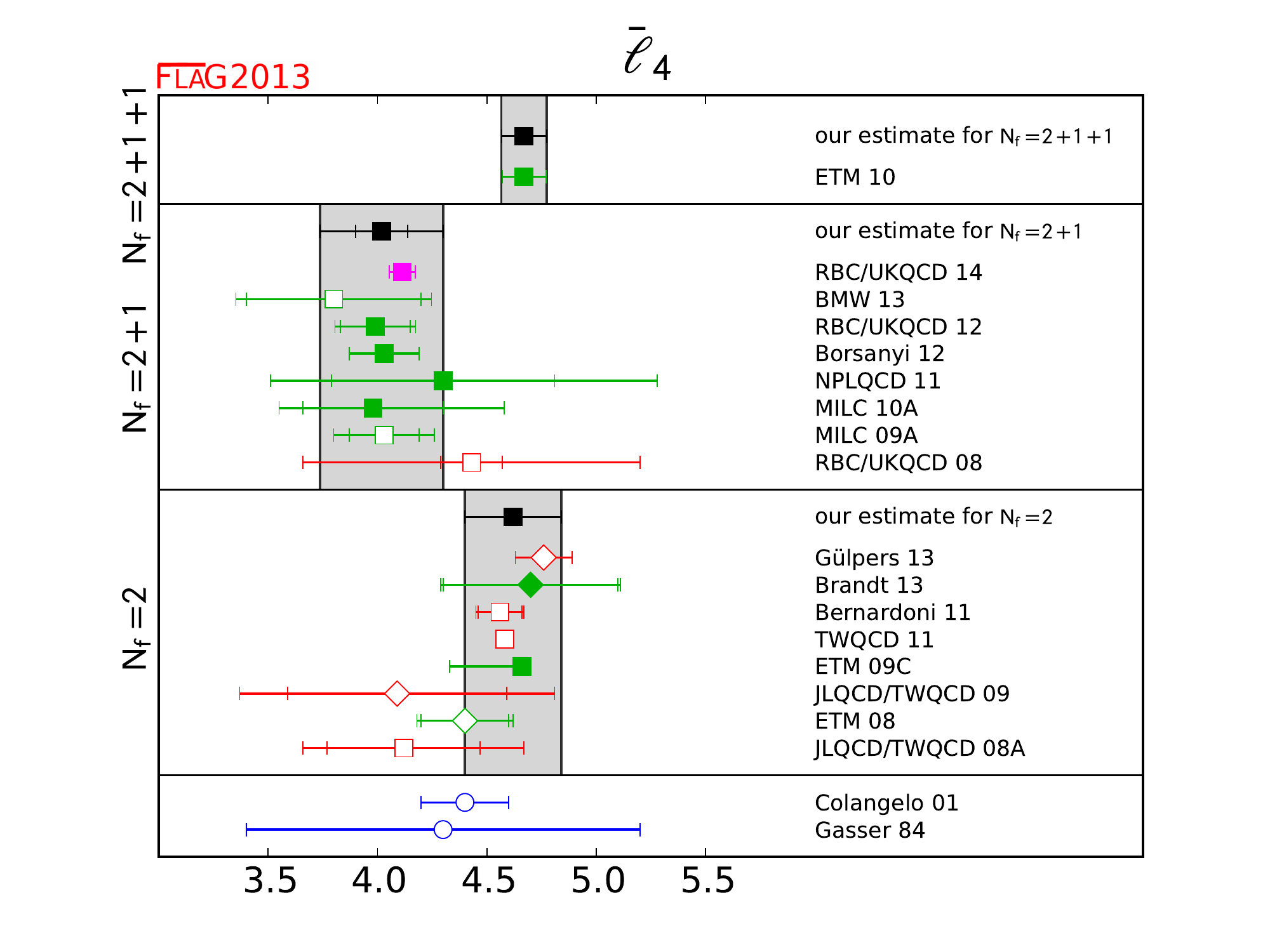}}\\
\vspace{-6mm}
\caption{\label{fig:su2-l34}
FLAG summary plots \cite{Aoki:2013ldr} from 2013 for the SU(2) LECs $\bar \ell_3$ (left) and $\bar \ell_4$ (right).
The magenta points, from RBC/UKQCD 14 \cite{Blum:2014tka}, have been added by me.
}
\vspace{-0.08in}
\end{figure}

The 2013 FLAG \cite{Aoki:2013ldr} summary plots   for the LECs $\bar \ell_3$ (left) and $\bar \ell_4$ (right) in SU(2) \chpt\ are
shown in \figref{su2-l34}, with the results from  RBC/UKQCD$\,$14 \cite{Blum:2014tka}  superposed in magenta.  
For $N_f=2+1$, the 2013 FLAG averages are
\begin{equation}
\bar\ell_3 = 3.05(99),\qquad \bar\ell_4= 4.02(28).
\eqn{su2-l34}
\end{equation}
The RBC/UKQCD$\,$14 results are
\begin{equation}
\bar\ell_3 = 2.73(13),\qquad \bar\ell_4= 4.113(59),
\eqn{su2-l34-RBC}
\end{equation}
again showing clearly the extent of recent progress.

The non-monotonic behavior with $N_f$ of $F_\pi/F$ seen in \figref{su2-Sigma-F} is seen again   in \figref{su2-l34} for 
$\bar\ell_4$. I suspect the same issue that affected $F_\pi/F$  is again at work here.  Indeed, D\"urr has noted
\cite{Durr:2014oba}, based on lattice data from \rcites{Borsanyi:2012,Durr:2013goa}, that $\bar \ell_4$ can be 
artificially raised
by $\sim\!10\%$ when the lower bound on the pion mass in the SU(2) \chpt\ fit is $\sim\! 270$ MeV, as it is in the ETM$\,$10 calculation, 
rather than close to physical.

\begin{figure}[t!]
  \centering
  {\includegraphics[width=0.48\linewidth]{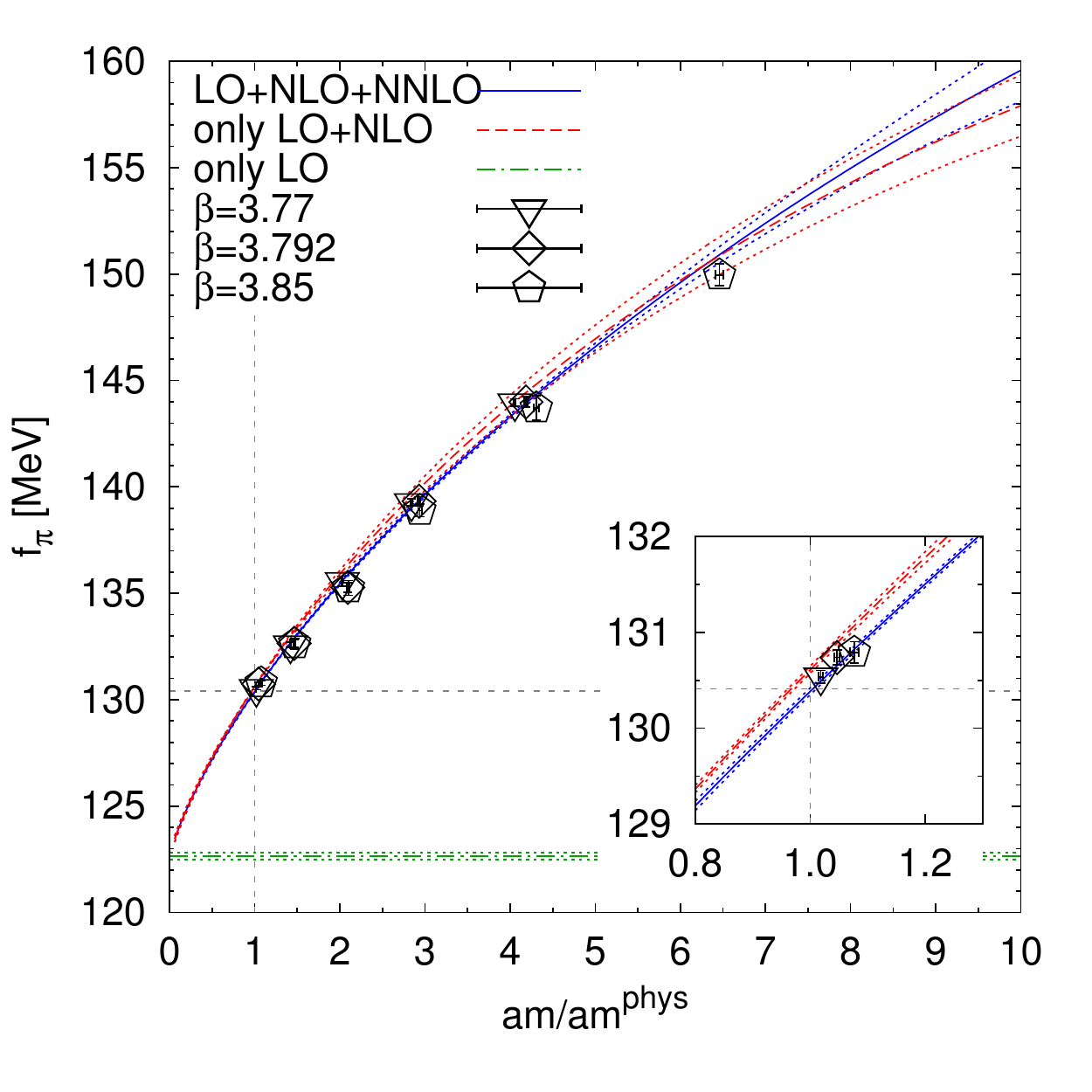}}
  \quad
  {\includegraphics[width=0.48\linewidth]{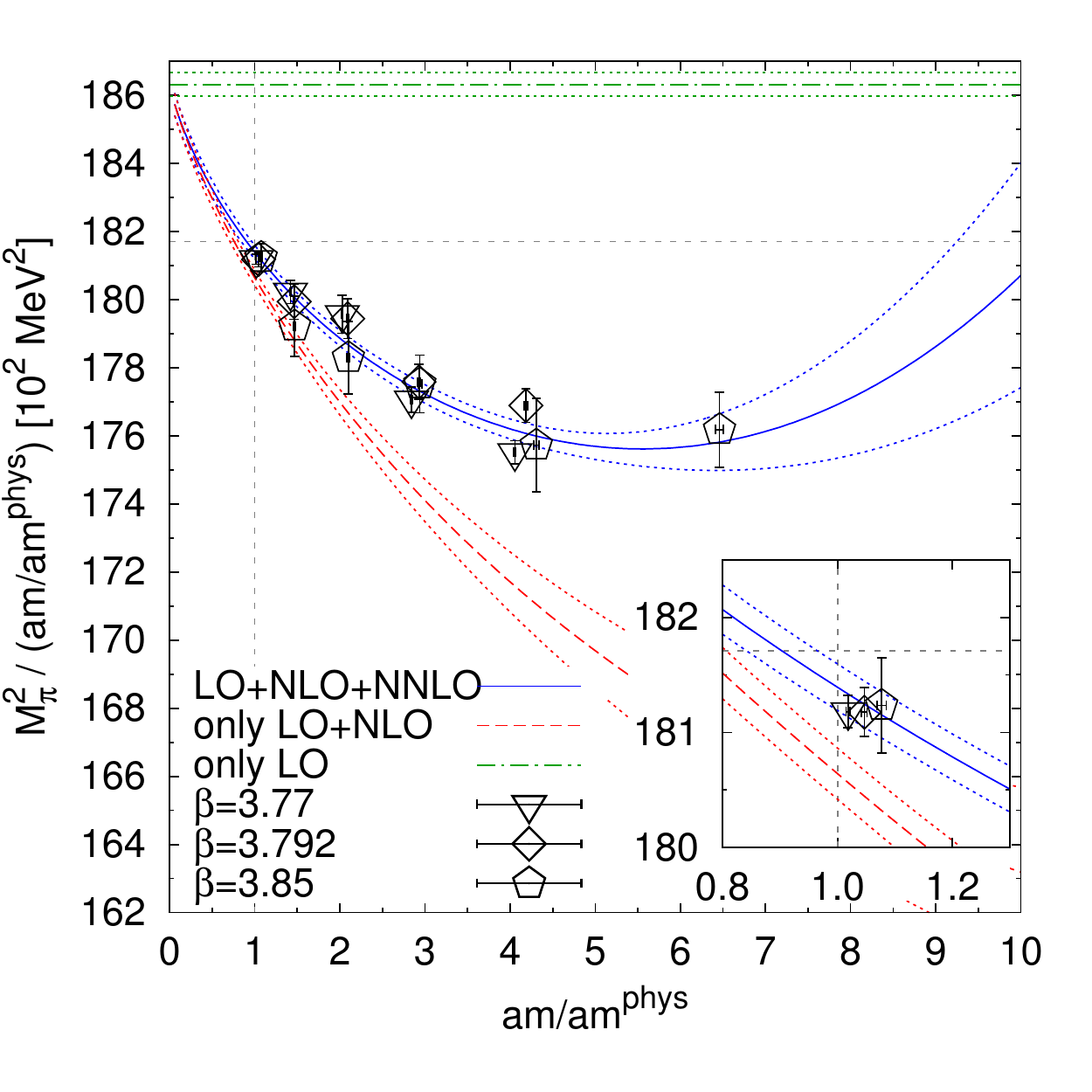}}\\
\vspace{-6mm}
\caption{\label{fig:su2-converge}
Convergence of SU(2) \chpt\ for the decay constant (left) and $M_\pi^2/\hat m$ (right) from \rcite{Borsanyi:2012}.
Points shown are at lattice spacings ranging from from $\approx\! 0.12$ fm to $\approx\! 0.10$ fm. There are $N_f=2+1$ flavors.
}
\vspace{-0.08in}
\end{figure}

A precise study of the ``convergence'' of SU(2) \chpt\ for the pseudoscalar decay constant and mass by Borsanyi 
{\it et al.}\ \cite{Borsanyi:2012} is shown in \figref{su2-converge}.  Their lattice data ranges from the physical value of 
$\hat m = (m_u+m_d)/2$ to about 6 times that value.
The convergence is good for  $f_\pi$:  LO+NLO (red) and LO+NLO+NNLO (blue) are almost on top of each other over the
full range of lattice data.   For $M_\pi^2/\hat m$, the convergence is less good: At the highest masses,
the NNLO contribution is $\sim\!60\%$ of the NLO contribution, but it drops to $\sim\!30\%$ by the time the
quark mass falls to $2\hat m^{\rm phys}$. Note, though, that both the NLO and the NNLO contributions  
are small compared to the LO term.  The picture seen here is in reasonable agreement with that seen in an earlier calculation by
the MILC Collaboration \cite{MILC-Lat09}; the lattice data there did not go down to physical quark masses, so required
an extrapolation to reach the region $\hat m \ltwid 2.5 \hat m^{\rm phys}$.

\begin{figure}[t!]
  \centering
  {\includegraphics[width=0.68\linewidth]{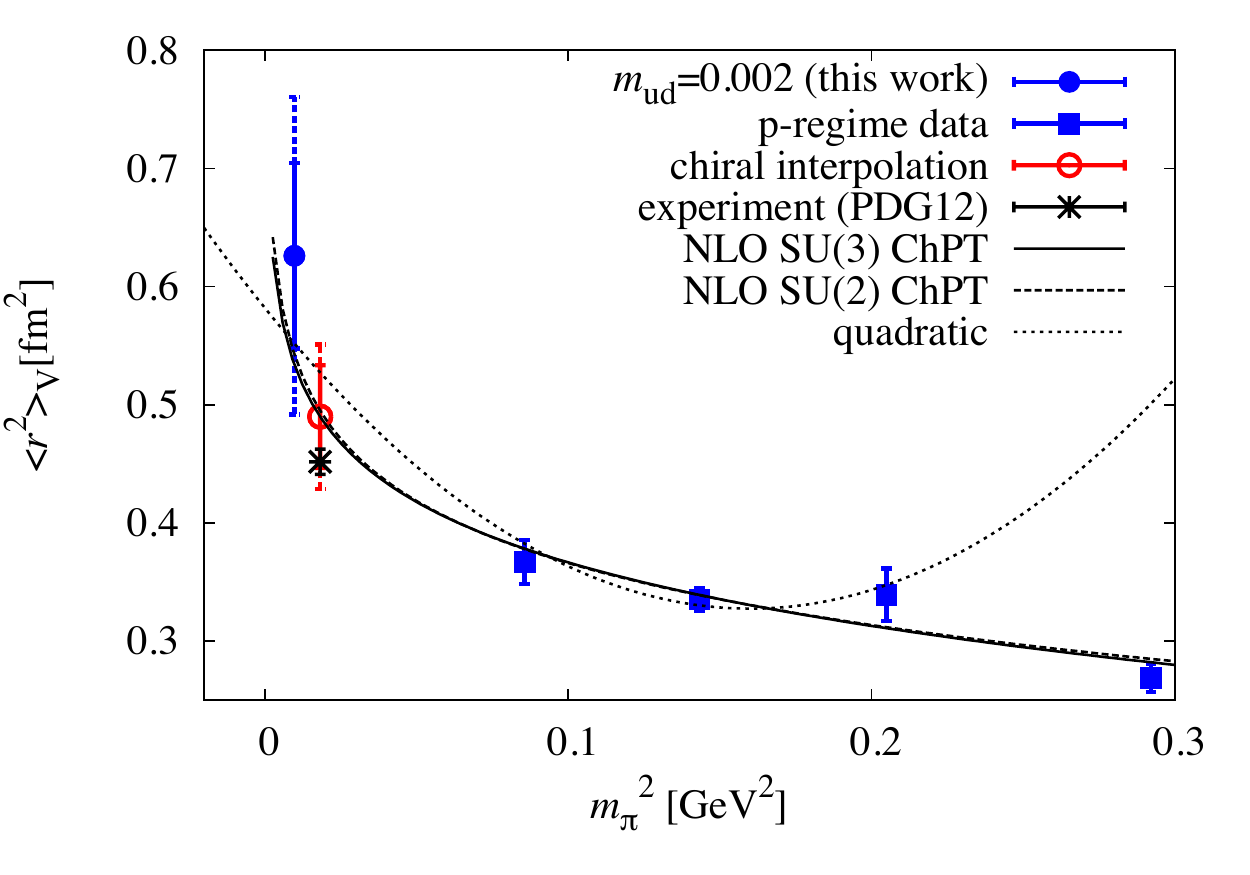}}\\
\vspace{-6mm}
\caption{\label{fig:epsilon-regime}
Interpolation of the mean-squared vector radius of the pion between the $\epsilon$-regime (blue circle) and the $p$-regime
(blue squares) by Fukaya \et\ \cite{Fukaya:2014}.  The red circle shows the interpolated value, which is compared to experiment (black burst).}
\vspace{-0.08in}
\end{figure}

I turn now to lattice determinations of the SU(2) LEC $\bar\ell_6$
and the pion mean-squared radius
$\langle r^2\rangle_V^\pi$ from the pion vector form factor. Form factors are calculated from lattice 3-point functions and are therefore 
considerably more difficult to determine than the LECs discussed earlier, which require only 
two-point Green's functions.   Indeed the FLAG 2013 average  \cite{Aoki:2013ldr}, $\bar\ell_6 = 15.1(1.2)$, comes
from only two calculations \cite{Brandt:2013dua,Frezzotti:2008dr}.
An very interesting recent paper by Fukaya \et\ \cite{Fukaya:2014} attempts to calculate $\langle r^2\rangle_V^\pi$
and $\bar\ell_6$ by interpolation between very small pion masses, in the $\epsilon$ regime of \chpt\ \cite{epsilon}, and larger
masses in the $p$ regime.
They use the overlap fermion formulation \cite{overlap} to keep exact chiral symmetry. \Figref{epsilon-regime} shows the
interpolation of $\langle r^2\rangle_V^\pi$ between the $\epsilon$-regime (blue circle) and $p$-regime points (blue squares).
Their fit gives:
\begin{eqnarray}
\langle r^2\rangle_V &=& 0.49 (4)(4) \; {\rm fm}^2\nonumber\\
F_\pi/F &=& 1.6(2)(3) \eqn{epsilon-regime} \\
\bar \ell_6 &=& 7.5(1.3)(1.5) \nonumber 
\end{eqnarray}
The interpolated value of the radius is in good agreement with experiment, but the lattice data seems to have
 too much curvature, resulting in quite low values for $F$ and $\bar\ell_6$.  \Rcite{Fukaya:2014} suggests that
 the problem may be the absence of NNLO terms in their analysis of the $p$-regime lattice data.  Another possible
 issue is the fact the small volumes they use (necessitated by the cost of overlap fermions) require novel methods
 in order to cope with potentially large finite-volume effects, such as the effects of fixed topology.

\begin{figure}[t!]
  \centering
  {\includegraphics[width=0.49\linewidth]{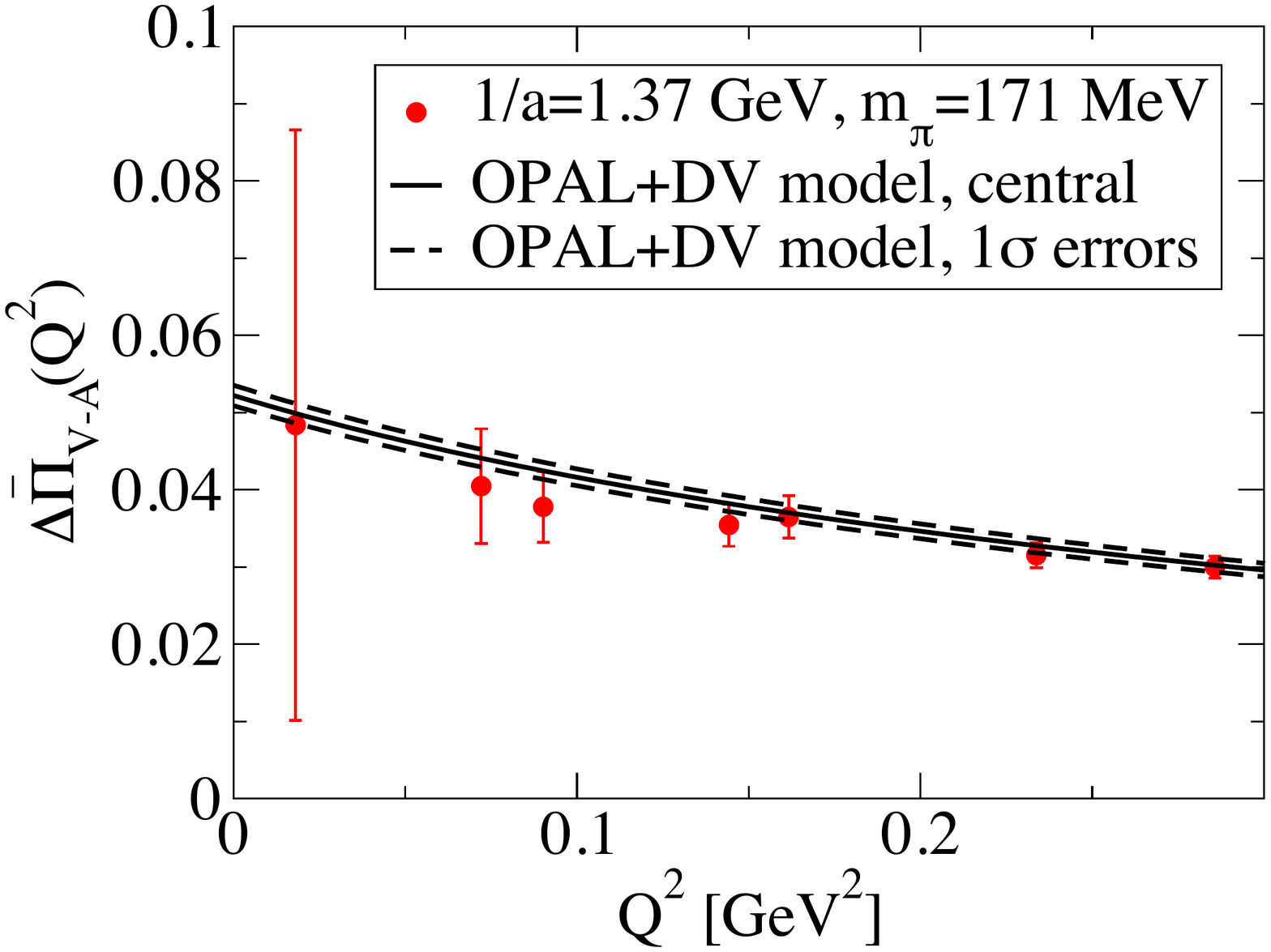}}
  \ 
  {\includegraphics[width=0.49\linewidth]{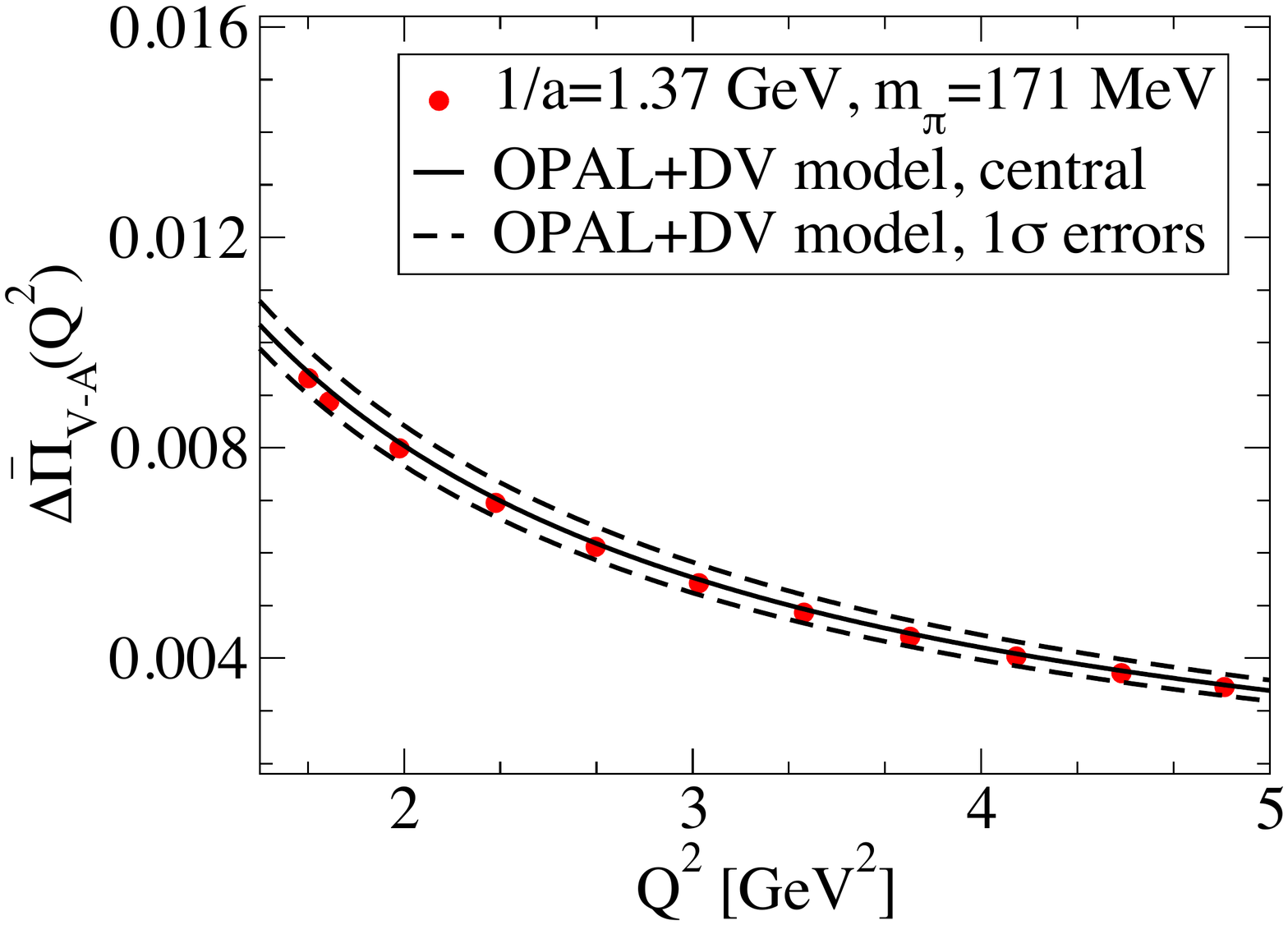}}\\
\vspace{-10mm}
\caption{\label{fig:L10} 
The spectral function $\Delta\bar \Pi_{V-A}$ from the $V-A$ $ud$ 
correlator as a function of $Q^2$ as determined by
experiment plus model (black lines) and by
lattice calculation (red points), from \rcite{Boyle:2014pja}.  The left-hand
plot shows the comparison at low $Q^2$; the right, at high $Q^2$.
}
\vspace{-0.08in}
\end{figure}

I would also like to call attention to recent work 
\cite{Boito:2012nt,Golterman:2014nua,Boyle:2014pja,Boito:2015fra}
on the SU(3)
LEC $L_{10}$, which appears in the correlators of vector and axial flavor
currents. \Figref{L10} shows a comparison  of lattice and continuum (experiment plus model)
determinations of  the pion-pole subtracted spectral function 
$\Delta\bar \Pi_{V-A}$ from the $V-A$ $ud$ correlator.  Note 
that the agreement is good for all
$Q^2$, but that the continuum results are more precise at low $Q^2$,
while the lattice results are more precise at high $Q^2$.  This means
that an approach using both lattice and continuum can do much better
than either alone. Furthermore, the lattice gives information on
the mass dependence of the correlators that is unavailable
from the continuum. In addition, these authors
use a chiral sum rule for the $ud-us$ correlator
in order to help constrain NNLO contributions that compete with that
of $L_{10}$.    The most recent result is \cite{Boito:2015fra}
\begin{equation}
L_{10}^r(m_\rho) = 0.00350(17),
\eqn{L10}
\end{equation}
which is  is fully controlled at NNLO and is the most precise determination of $L_{10}^r$ to date. The approach in these
papers is especially appealing because it uses information from
all available sources (experiment, continuum phenomenology, lattice)
in order to extract the result.

\subsection{A Brief Discussion of Nucleons}
\label{sec:nucleons}

\begin{figure}[t!]
  \centering
  {\includegraphics[width=0.70\linewidth]{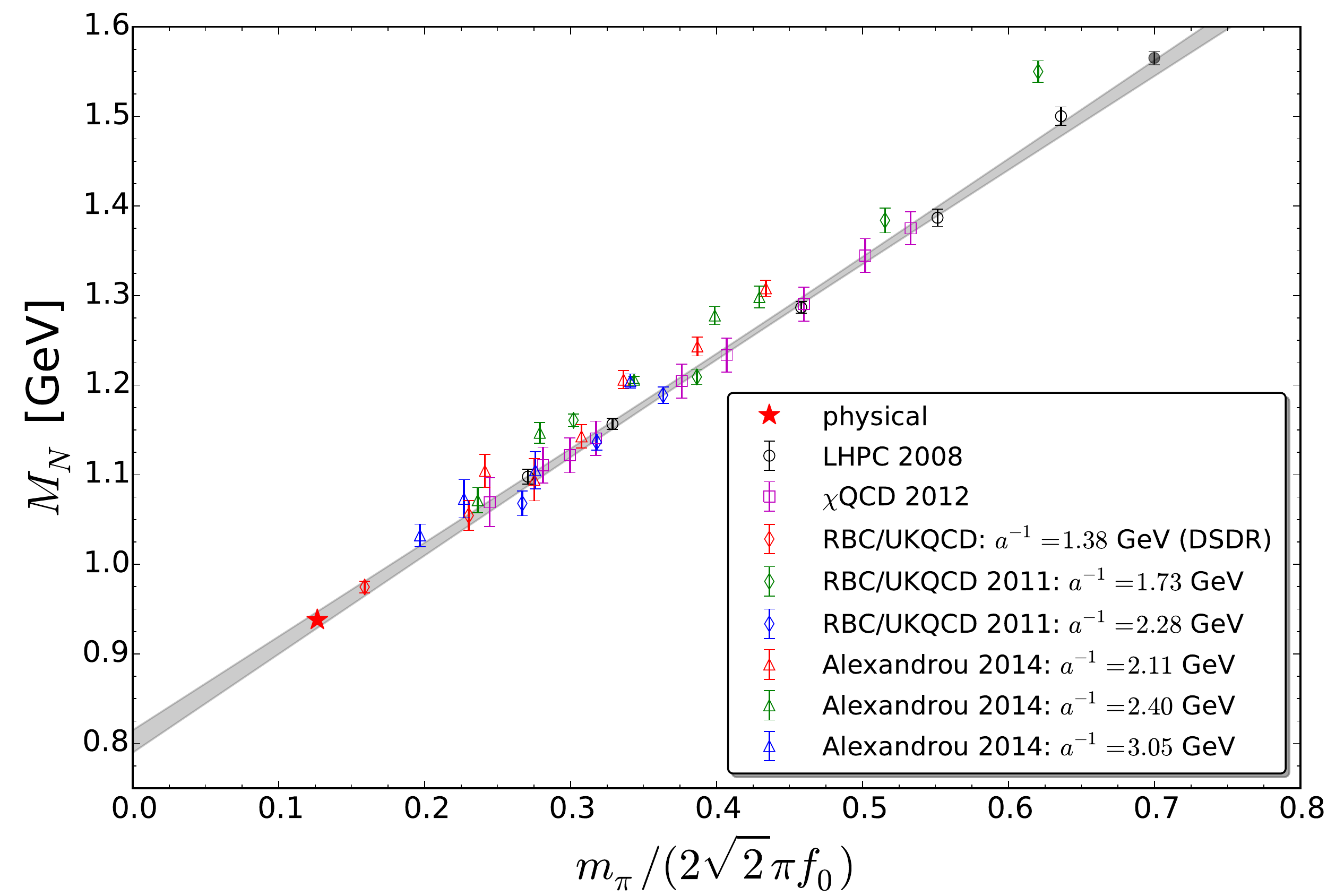}}\\
  \vspace{-2mm}
\caption{\label{fig:ruler} 
The ``ruler plot'' \cite{Walker-Loud:2014iea,WalkerLoud:2008bp}
 for the mass of the nucleon as a function of $m_\pi$, with some updates as
described in the text.  The gray band is a linear fit to the LHPC points only \cite{WalkerLoud:2008bp}.
}
\end{figure}

The dependence of the nucleon mass on the pion mass remains a puzzle.  As emphasized by Walker-Loud and collaborators
\cite{WalkerLoud:2008bp,Walker-Loud:2014iea}, the dependence is quite linear up to rather high pion masses ($\sim\!\!700$ 
MeV or higher).  \Figref{ruler}  shows a recent version of the so-called ``ruler-plot'' \cite{Walker-Loud:2014iea}, where
I have updated the Alexandrou \et\ and RBC/UKQCD points using their published results \cite{ETM:2014,Aoki:2010dy} where
possible; for the RBC/UKQCD $a^{-1}=1.38$ GeV DSDR nucleon masses, I used updated but still unpublished lattice data
\cite{Blum}.  While the newer lattice data appears to favor a slightly greater slope than the original fit to the LHPC points only,
the linearity still seems clear.
This behavior with $m_\pi$ has no obvious physical explanation. In heavy-baryon \chpt\  \cite{heavy-baryon} there are $m_\pi^2$ and $m_\pi^3$
terms at NLO, and an $m_\pi^4\ln{m_\pi^2}$ term at NNLO, but no linear term at all.  Other versions of baryon \chpt\ 
(Lorentz invariant \cite{Becher:1999he}, with  explicit $\Delta$ degrees of freedom \cite{Jenkins:1991es}, {\it etc.}) do not change the picture. Nor is the picture changed by the inclusion of lattice data from groups not represented in \figref{ruler}.
I call the reader's attention to Fig.~5 in \rcite{ETM-Lat14-baryon}. which shows that results for $m_N$ as a function
of $m_\pi$ from a large selection of groups are in good agreement.

\begin{figure}[t!]
  \centering
  {\includegraphics[width=0.70\linewidth]{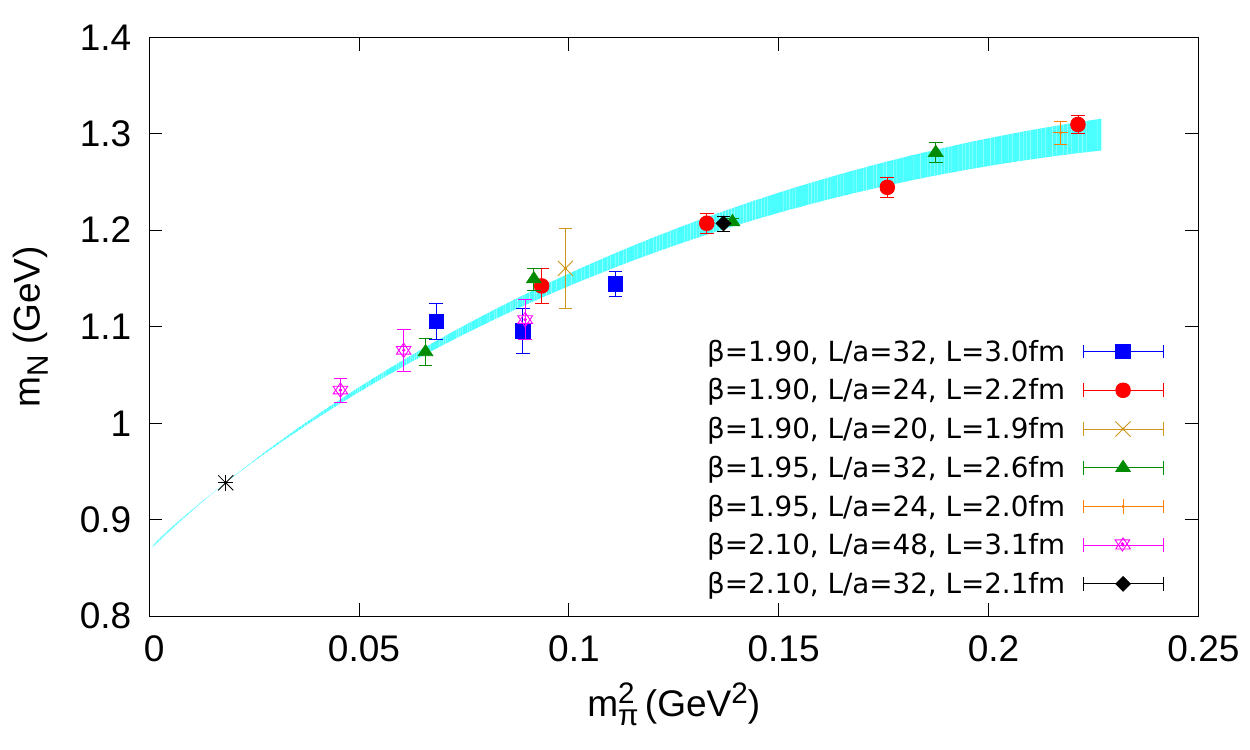}}\\
\vspace{-4mm}
\caption{\label{fig:ETM_mN} 
A chiral fit to 
{\protect\eq{HB_NLO}}
for the mass of the nucleon as a function of $m^2_\pi$, by 
C.\ Alexandrou {\it et al.}\ \cite{ETM:2014}. The fit is constrained to pass through the physical point, indicated by the black 
burst.}
\vspace{-0.05in}
\end{figure}

It is true, however, that \chpt-inspired fits of the lattice data are possible.  For example, \figref{ETM_mN} shows a fit from
\rcite{ETM:2014} to the heavy-baryon \chpt\ form at NLO:
\begin{equation}
m_N=m_N^{(0)}-4c_1m^2_\pi - \frac{3 g_A^2}{32\pi F_\pi^2}m_\pi^3.
\eqn{HB_NLO}
\end{equation}
The horizontal axis is now $m_\pi^2$, not $m_\pi$, so there is clear curvature.
The value used for the parameter $c_1$ is $\approx \! -1.14\; {\rm GeV}^{-1}$,  consistent with phenomenological expectations 
\cite{Kubis}, and the fit is good.  However, the problem is that the next order term, whose coefficient is known from 
pion-nucleon Roy-Steiner equations, would ruin the fit at fairly low values of $m_\pi$, not far above the physical
point \cite{Kubis}.  Indeed, to get the ultimately quite linear behavior seen in \figref{ruler} would require what seems like a 
``conspiracy'' among higher order terms.  In his talk here, S.\ Beane \cite{Beane} has pointed to the existence
of several additional such
conspiracies in chiral dynamics. The physical reasons for these conspiracies are not
known.

\subsection{SU(3) and the Three-flavor Chiral Limit: Preliminary Results} 
 \label{sec:chiral-limit}
 
Studying the convergence and the chiral limit in mesonic \chpt\ is a much more difficult job on the lattice for SU(3) than for SU(2).
We know   $f_K/f_\pi\approx1.2$, but $f_K=f_\pi$ at LO in SU(3) \chpt, which implies that there 
are $\sim\!20\%$ corrections at NLO at the physical strange-quark mass $m_s$.
So even if SU(3) \chpt\ ``converges well,'' one would expect  $\sim\!4\%$ NNLO corrections. 
Because lattice data for quantities such as meson masses or decay constants has sub-percent statistical 
errors, good fits 
require terms of still higher order.
But the chiral logs are not known at N${}^3$LO, so the fits become {\it ad hoc} at this order.
In practice this is fine if the goal is simply interpolating around $m_s$, as is, for example, needed for
calculations of $f_K$ and $f_\pi$; the extrapolation in light-quark ($u$, $d$) mass that may be required is
insensitive to such higher order terms.  But for more fundamental issues involving the structure of
\chpt, the {\it ad hoc} nature of the fits -- not to mention the fact that the expansion is asymptotic, which
should discourage us from going to arbitrarily high order at fixed mass --- may lead to an unacceptable
ambiguity in interpreting the results.  Examples of such more fundamental questions are: (1) assessment of 
the rate of convergence, (2) the nature of the 3-flavor chiral limit, which would involve extrapolation in the 
simulated value of the strange mass, $m'_s$, from the physical value  $m'_s=m_s$ to $m'_s=0$, and (3) 
finding the values of (some of the) LECs.  Those LECs associated with terms involving the strange sea-quark mass,
such as $L_4$ and $L_6$, are likely to be sensitive to the higher order terms since $m_s$ is large.

An additional problem is the following:  The coupling constant of SU(3) \chpt\ is $1/(16\pi^2 f_0^2)$, where $f_0$ is the decay   
constant in the 3-flavor chiral limit.   However, for the highest order in a systematic expansion (NNLO in practice), it is 
consistent (and even sensible) to replace $f_0$ by a physical decay constant,  such as $f_\pi$  or even  $f_K$  (if 
the data to be fit runs up to the kaon mass).  This makes a big difference in the size of the NNLO terms, whose coefficient
goes like $1/f_0^4$.

My conclusion from the above discussion is that reliable control of SU(3) \chpt\ is only possible for the simulated 
strange-quark mass $m_s'$ chosen to be less than its physical value  $m_s$.  The MILC Collaboration has generated
a useful set of ensembles with the asqtad staggered action (circa 2009 \cite{Bazavov:2009bb}) that have
$0.1m_s \le m'_s \le 0.6 m_s$.   SU(3) \chpt\ on these ensembles was studied in 2010 \cite{MILC-Lat10}, but the
results were inconclusive.  Here, I present a second (but still preliminary) look at our lattice data.

The fits I  describe are all systematic fits through NNLO, in that they include the complete (partially
quenched) chiral logarithms \cite{Bijnens-PQ}  as well as the
corresponding analytic terms.  The staggered discretization effects on the NNLO terms are omitted, however, because they have not yet been calculated. Nevertheless, the resulting chiral forms are systematic expansions through NNLO 
as long as the discretization 
effects at NNLO are small compared to the included mass-dependent terms; this can be arranged by cutting the data to
remove coarse lattice spacings where the discretization terms are not negligible. In practice, ensembles with $a\ge0.12$ fm 
need to be dropped. There are then two remaining spacings, $a\approx 0.09$ fm and $a\approx 0.06$ fm, in the 
analysis.
 
\begin{figure}[t!]
  \centering
  {\includegraphics[width=0.49\linewidth]
  {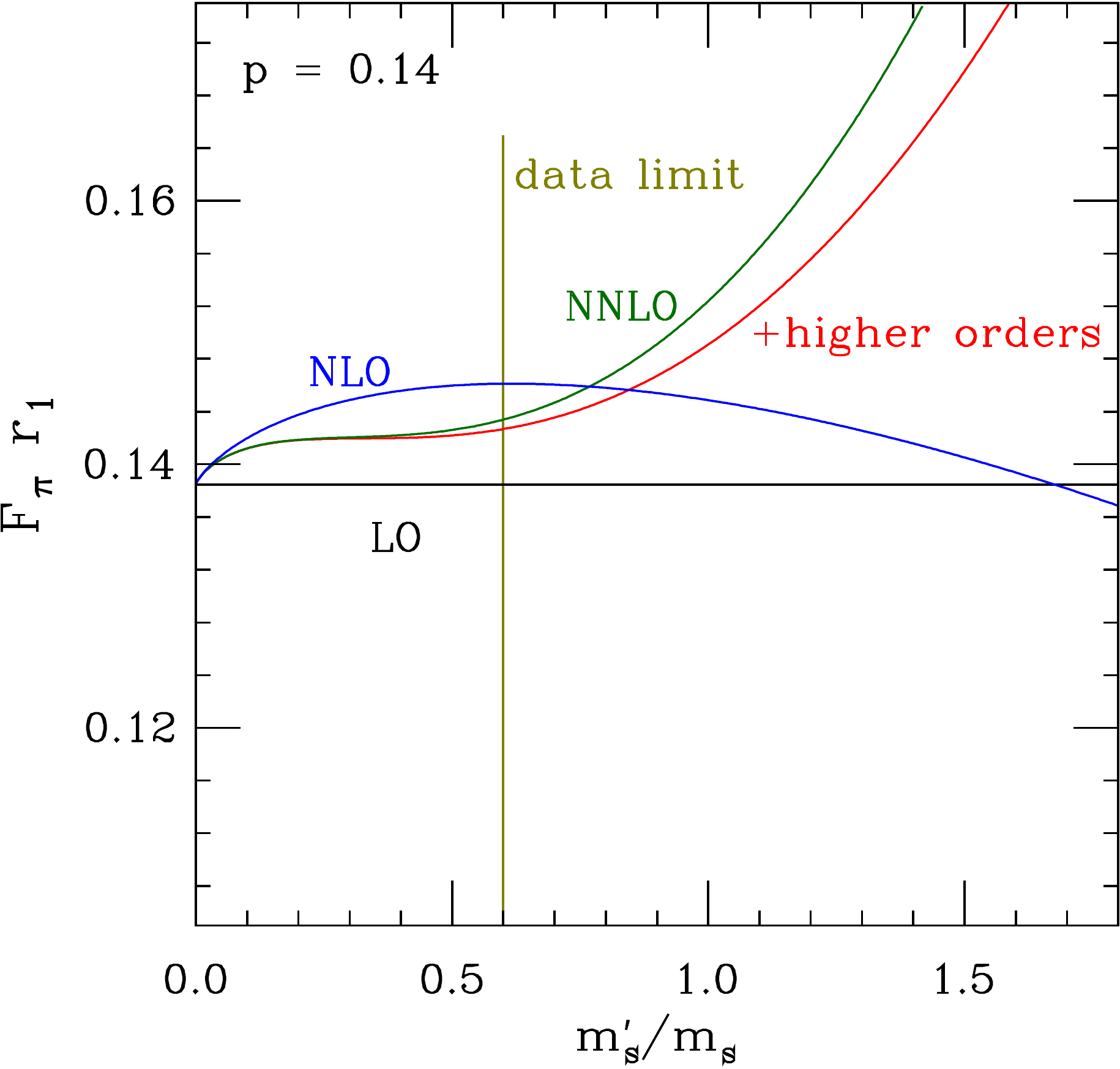}}
  \ 
  {\includegraphics[width=0.49\linewidth]
  {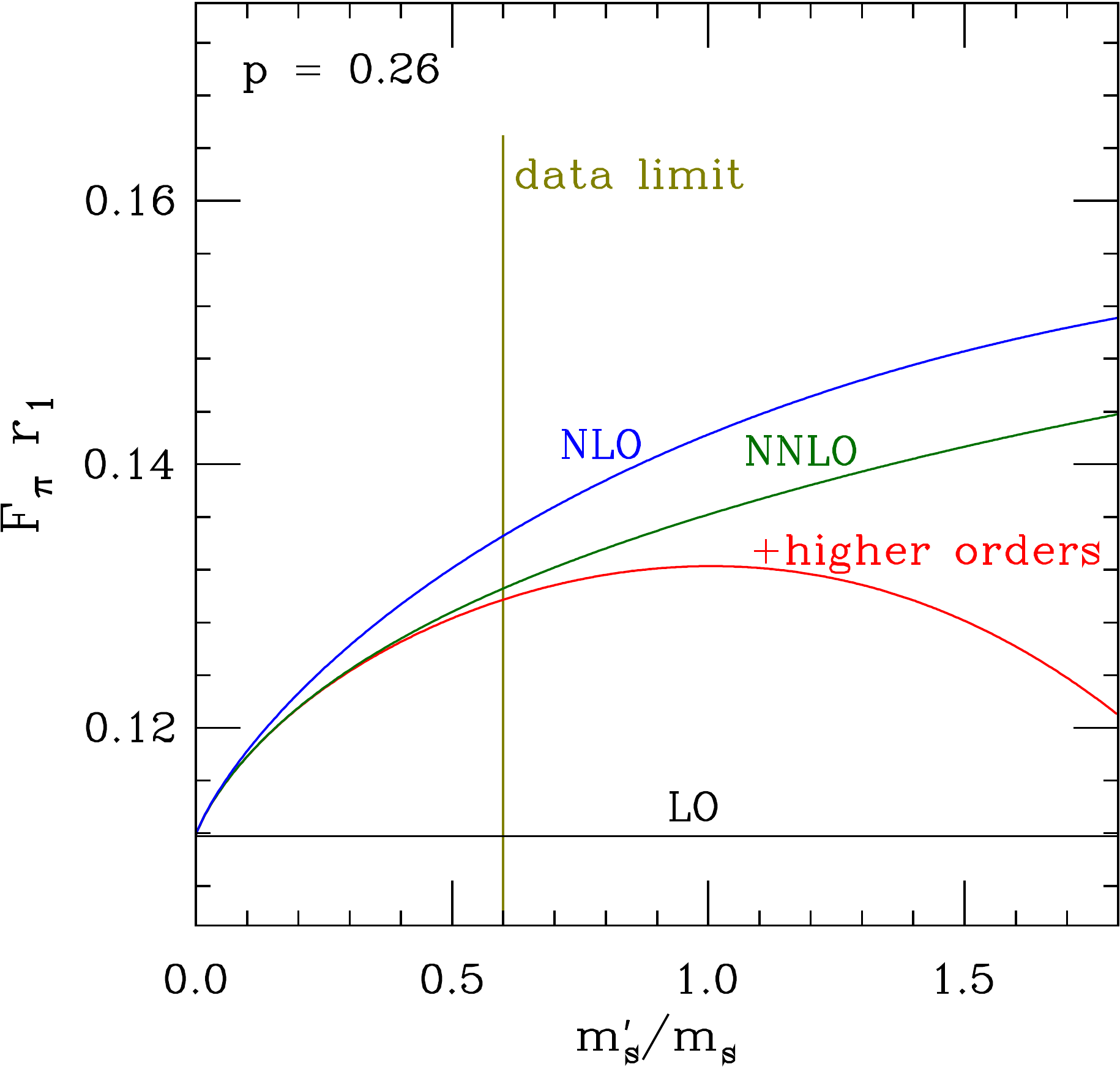}}\\
\vspace{-2mm}
\caption{\label{fig:SU3-fpi-AB} 
Behavior of the decay constant determined from SU(3) NNLO fits (left: type A; right: type B) 
to decay constants and masses, extrapolated to the continuum, and plotted as a function of the 
simulated strange sea mass over the
physical strange mass, $m'_s/m_s$, with the $u,d$ quark mass $\hat m$ extrapolated to the chiral limit.  
Lines labeled LO, NLO, and NNLO show the contributions up to and including the indicated 
order in \chpt.  Lines labeled ``higher order'' show the 
effect of adding N${}^3$LO analytic terms to the fit, while holding the LO, NLO, and NNLO 
terms fixed; note that little effect is seen until above the 
limit of the lattice data at $m'_s/m_s=0.6$.
These fits to MILC asqtad lattice data are very similar to those described in \rcite{MILC-Lat10}.
}
\vspace{-0.08in}
\end{figure}

Let us call the value of the decay constant chosen to appear in front of the NNLO terms $f_{NNLO}$.  As might be
expected on physical grounds, fits with $f_{NNLO}=f_0$ are poor, with low $p$-values, $p\le0.01$. Acceptable
fits with $f_{NNLO}\approx f_\pi$ are possible.  Two different ways of setting   $f_{NNLO}\approx f_\pi$ have been
tried:   ``Type-A'' fits set $f_{NNLO}=\tau f_0$, where $f_0$, the decay constant in the chiral limit, is a fit parameter, 
and $\tau$ is a fixed number, determined iteratively to make  
$f_{NNLO}\approx f_\pi^{\rm expt}$, the experimental pion decay constant.  ``Type-B'' fits simply fix 
$f_{NNLO}=f_\pi^{\rm expt}$.  These two fit versions give disconcertingly
different pictures of the 3-flavor chiral limit and the convergence of SU(3) \chpt, as seen in \figref{SU3-fpi-AB}.
The type-A fit results in a large value of the LO contribution $f_0$ and small corrections at NLO, whereas the type-B
fit gives a small value of $f_0$, and correspondingly large corrections at NLO.  Indeed the latter fit
gives $f_\pi/f_0=1.26(4)$, and, for the condensates, $|\langle \bar u u\rangle_2| / |\langle \bar u u\rangle_3| = 1.59(13)$, where the subscripts
2 and 3 denote the 2- and 3-flavor chiral limits.  This strong suppression of the 3-flavor chiral-limit quantities, compared
to the 2-flavor values, is suggestive of the ``paramagnetic effect'' \cite{DescotesGenon:1999uh}.  
On the other hand, the type-A fit has
only small suppression of the 3-flavor chiral-limit quantities:
$f_\pi/f_0=1.09(2)$ and $|\langle \bar u u\rangle_2| / |\langle \bar u u\rangle_3| = 1.18(8)$.   

The reason for the large differences between the different versions of the chiral fit is not hard to discover.  The 
lattice
data favors small contributions at NNLO; in other words, it wants $f_{NNLO}$ to be large.  This can be 
deduced from the 
 fact that fits with $f_{NNLO}=f_0$ are poor.  In type-A fits, $f_{NNLO}$ can be
raised by increasing $f_0$ for fixed $\tau$.    In type-B fits $f_{NNLO}=f_\pi^{\rm expt}$, independent of the 
value of $f_0$.  
The fit can then further improve by redistributing the amount of LO and NLO contributions: smaller LO and 
larger NLO ones
are preferred.  Note that the $p$-value of the type-B fit (0.26) is somewhat higher than that of the type-A one 
(0.14), but both fits are acceptable. The large discrepancy between the
results of the two types of fit, with no compelling reason to reject either, is what deterred us from 
publishing a final version
of the analysis described in \rcite{MILC-Lat10}.

\begin{figure}[t!]
  \centering
{\includegraphics[width=0.55\linewidth]{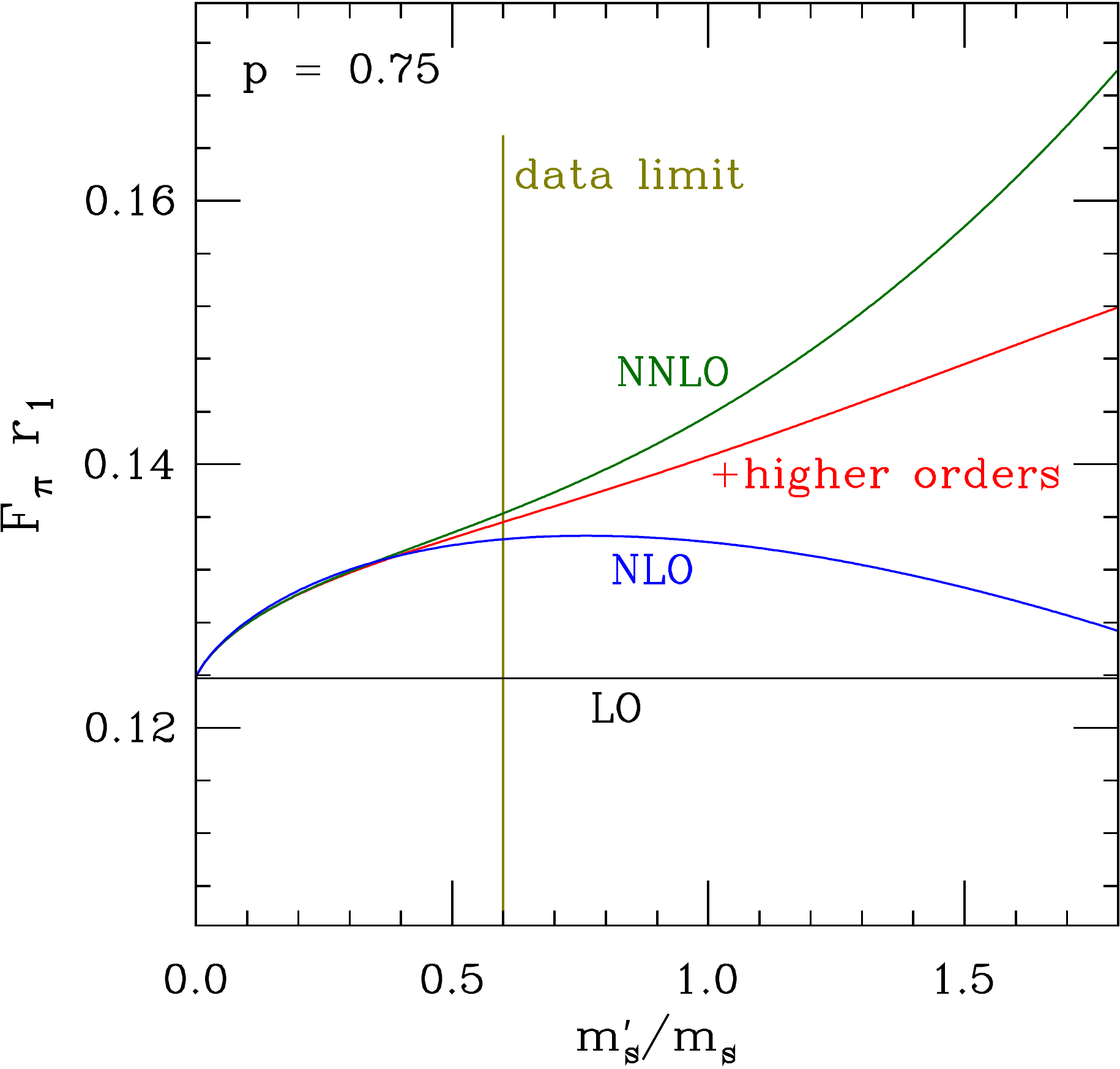}}
\vspace{-3mm}
\caption{\label{fig:SU3-fpi-fK} 
Same as the plot shown in
{\protect\figref{SU3-fpi-AB}} (left) but with the choice $f_{NNLO}\approx f_K$.
}
\end{figure}

Recently, we have revisited this situation. Because the maximum value of $m'_s$ 
is $0.6m_s$, the highest meson masses considered are slightly 
larger than $m_K$. This, coupled with the fact that the lattice
data prefers high values of $f_{NNLO}$, suggests trying fits with 
$f_{NNLO}$ as large as $\sim\!f_K$.   \Figref{SU3-fpi-fK} shows one such fit.  It 
happens to be a type-A fit, but once $f_{NNLO}$ gets near $f_K$ the big difference between type-A and type-B fits goes away.  
When $f_{NNLO}$ 
is large enough, the NNLO terms no longer control the preferred size of $f_0$ 
in the type-A case, and $f_0$ gets adjusted similarly in both
A and B fits.  Note that the $p$-value of the fit in \figref{SU3-fpi-fK} is now 0.75, 
much better than either of the fits with  $f_{NNLO}\approx f_\pi$
shown in \figref{SU3-fpi-AB}.  When $f_{NNLO}$ is made a free parameter, it rises slightly 
further,  to about $1.1 f_K$.  However, the
picture in \figref{SU3-fpi-fK} barely changes, and  in fact the $p$ value of the fit decreases slightly.  

Taken at face value, this suggests using the results of the fit in \figref{SU3-fpi-fK} as central values, and taking the two extremes seen in \figref{SU3-fpi-AB} as defining the systematic errors.  This gives (preliminarily):
\begin{equation}
\eqn{chiral-limit-results}
\frac{f_\pi}{f_0}=1.17(4)(9);\hspace{13mm} \frac{|\langle \bar u u\rangle_2|}{|\langle \bar u u\rangle_3|} = 1.34(10)({}^{+25}_{-16}).
\end{equation}
I would characterize  \figref{SU3-fpi-fK} as showing good convergence of SU(3) \chpt\ up until $m'_s\sim\!0.7m_s$, but significantly poorer 
convergence, or even a ``breakdown,'' as $m_s$ is approached or surpassed.  

A concern with the present analysis is that the sensitivity to the value of  $f_{NNLO}$  
seems to indicate that N${}^3$LO effects are not negligible. 
However, it may instead be that the omission of discretization errors in our NNLO form is pushing the fit to 
suppress these terms by making $f_{NNLO}$ large, and that the sensitivity to $f_{NNLO}$ is a lattice artifact.  
To get a better handle on these questions, one needs lattice data with smaller discretization errors and smaller $m_s'$.  Such an improved analysis,
using the highly improved MILC HISQ ensembles \cite{Bazavov:2012xda}, is in progress.   Higher order staggered \chpt\ calculations would also 
help in clarifying the issues; I understand that J.\ Bijnens and collaborators are working on such calculations.   

As is known from both continuum analysis \cite{Ecker:2013pba} and lattice simulations, the value of $f_0$ is strongly anti-correlated with that of the LEC 
$L_4$.  At the physical point,  $L_4$ is multiplied by $m_s$, and because $m_s$ is large, a smaller value of $f_0$ can be approximately
compensated, for many quantities, by a larger value for $L_4$.    Thus \rcite{Ecker:2013pba} suggests trying to extract the appropriate linear
combination of $f_0$ and $L_4$, which should be significantly better determined than either quantity alone.  We will do this in the next version
of the MILC analysis.

\section{Prospects for the Future}
\label{sec:prospects}

\subsection{\chpt\  in service to lattice calculations}   
Now that ensembles with near-physical quark masses are feasible,  chiral fitting and extrapolation is less important than it used to be.
Still, \chpt, allows us to include other ensembles with higher quark masses, which often have smaller statistical errors.
In addition, \chpt\  helps control finite volume effects, and that role for \chpt\ is unlikely to change in the future.

The modern prevalence of highly improved actions and small lattice spacings means that lattice artifacts can become
so small that simple analytic expansions in the lattice spacing $a$ are good enough.
Furthermore, with improved actions, the nominally-leading discretization terms that are included 
in \chpt\ may have their coefficients reduced so much that their effects become comparable to those of higher order terms in $a$, which 
are difficult to include in \chpt.  
In many future lattice computations, it may therefore be unproductive  to include discretization errors in \chpt.     
The condition for the calculation of physical quantities 
not to need lattice \chpt\  is that physical $m^2_\pi$ be much larger than the symmetry-breaking 
lattice artifacts. 
For staggered HISQ quarks, this condition is now well satisfied for $a \le 0.06$ fm  (\ie for some of the HISQ ensembles 
MILC uses, but not yet for all the ensembles used in controlled calculations).
On the other hand, for staggered calculations of chiral-limit quantities such as $f_0$, 
staggered \chpt\  will continue to be needed:  As one approaches the chiral limit,  quark-mass effects eventually
become smaller than staggered taste-breaking effects, and the latter must be included in performing the extrapolation.

\subsection{The  ``payback'' by lattice QCD to \chpt} 
The payback to \chpt\ will continue to improve.
It is straightforward to reduce errors on the simple mass-dependent SU(2) LECs,  $\bar\ell_3$ and  $\bar\ell_4$, and this will be
done in the near future.
The current small discrepancies between groups should go away as more and more calculations use ensembles with quark
 masses down to (or even below) the physical light quark masses.  Other LECs, including those at NNLO, can also be extracted as the precision 
 improves.
 
In SU(3) \chpt, the generation of dedicated ensembles with  $m'_s < m_s$  seems necessary in general to get good control.
This is especially true for 3-flavor chiral-limit quantities $f_0$ and the chiral condensate, 
and probably for the NLO sea-quark-dependent LECs  $L_4$  and $L_6$.
Because such ensembles (especially if $m'_s\ll m_s$ )  are not particularly useful for most other lattice QCD calculations (\eg flavor physics),
lattice groups will only generate them if there is strong demand from the \chpt\  community.

Of course, there is a huge QCD world out there, and  in this talk I  only have been able to describe the lattice calculation of a small number of rather simple
quantities. Lattice QCD is exploring more and more issues from first principles, and with control over systematics:
form factors, scattering amplitudes, baryons and light nuclei, hadronic weak decays, electromagnetic and isospin-violating effects, hadronic 
contributions to $(g-2)_\mu$.  So there will be many paybacks to come!           
\bigskip

{\bf Acknowledgments:} This work was supported in part by the U.S. Department of Energy under 
Grant DE-FG02-91ER-40628.  I thank J.\ Bijnens, T.\ Blum, U.\ Heller, P.\ Masjuan, B.\ Tiburzi, and A.\ 
Walker-Loud for useful communications and discussion, and M.\ Golterman, S.\ Gottlieb, U.\ Heller, J.\ Komijani, and 
R.\ Sugar for carefully reading this manuscript and for many helpful suggestions.

\end{document}